\documentclass[prd,aps,notitlepage,twocolumn,longbibliography,nofootinbib]{revtex4-1} 
\usepackage{amsfonts,amssymb,amsmath,mathrsfs,graphicx,xcolor,bm,mathtools,physics,orcidlink,url,revsymb}
\let\standardnot\not
\let\standardwidehat\widehat
\let\standardwidetilde\widetilde
\usepackage[mathx]{mathabx}
\let\not\standardnot
\let\widehat\standardwidehat
\let\widetilde\standardwidetilde
\definecolor{ultramarine}{rgb}{0.07, 0.04, 0.56}
\definecolor{cadmiumgreen}{rgb}{0.0, 0.42, 0.24}
\definecolor{indigo(dye)}{rgb}{0.0, 0.25, 0.42}
\usepackage{hyperref}
\hypersetup{
colorlinks=true,
citecolor=ultramarine,
linkcolor=cadmiumgreen,
urlcolor=indigo(dye),
}

\newcommand{\be}{\begin{equation}}
\newcommand{\ee}{\end{equation}}

\renewcommand{\dd}{\mathrm d}
\newcommand{\ii}{\mathrm i}
\newcommand{\Id}{\mathbf 1}

\numberwithin{equation}{section}
\renewcommand{\theequation}{\arabic{section}.\arabic{equation}}

\begin{document}
\title{Charge-regular formulation and radiative amplitudes of\\
Kerr--Newman perturbations}
\author{Hayato Motohashi\,\orcidlink{0000-0002-4330-7024}}
\affiliation{Department of Physics, Tokyo Metropolitan University, 1-1 Minami-Osawa, Hachioji, Tokyo 192-0397, Japan}

\begin{abstract}
We establish a charge-regular formulation of separated Kerr--Newman perturbations and an explicit map from master solutions to physical radiative fields.
We derive regular angular and radial first-order systems and their second-order reductions, with finite-charge deformations of the spin $\pm1$ and $\pm2$ Teukolsky equations.
The field and boundary normalizations recover the standard Kerr conventions while accounting for the nonuniform zero-charge limit at infinity.
For physical Einstein--Maxwell perturbations with the usual peeling behavior, we obtain explicit outgoing gravitational and electromagnetic amplitudes and energy fluxes at nonzero real frequency.
The resulting finite-charge radiation-composition ratios depend only on the regular angular functions, independently of the radial excitation.
\end{abstract}
\maketitle

\tableofcontents

\section{Introduction}\label{sec:intro}

Gravitational and electromagnetic perturbations of Kerr are described by the separated Teukolsky equations~\cite{Teukolsky:1972my,Teukolsky:1973ha}.
Their connection to radiative boundary conditions and energy fluxes made the separated equations useful for stability and scattering calculations~\cite{Press:1973zz,Teukolsky:1974yv}.
A neutral massless scalar test field on Kerr--Newman also admits angular and radial separation~\cite{Carter:1968ks,Boonserm:2014rma}.
The coupled gravitational and electromagnetic perturbations present a different problem, because the background electric field couples the two sectors already at linear order.
The nonrotating Reissner--Nordstr\"om (RN) problem admits a gauge-invariant reduction~\cite{Moncrief:1975sb}, with decoupled channels whose mixtures determine the gravitational and electromagnetic radiation~\cite{Chandrasekhar:1985kt,OuldElHadj:2023anu}.
For simultaneous rotation and charge, exact separation of this coupled radiative system remained unresolved until the recent construction discussed below.

Earlier Kerr--Newman studies addressed the coupled problem through complementary approaches.
Perturbative calculations obtained mode spectra at first order in rotation~\cite{Pani:2013ija,Pani:2013wsa} and at small charge without a slow-rotation expansion~\cite{Mark:2014aja}.
Dias, Godazgar, and Santos computed quasinormal modes from coupled equations in radius and polar angle~\cite{Dias:2015wqa,Dias:2022oqm}; their extended study also gives a gauge-invariant formulation of the radiative fields.
Giorgi derived coupled Teukolsky and Regge--Wheeler systems suitable for physical-space analysis~\cite{Giorgi:2020ujd}.
These results establish that a regular perturbation formulation and numerical access to its spectrum do not require an exact separation.
A separated description with fixed physical normalization nevertheless offers a different starting point for connection and radiation calculations.

Hintz recently separated the radiative Kerr--Newman system using a matrix integrating factor for its first-order equations and proved real-axis mode stability for the resulting radial problem~\cite{Hintz:2026jiw}.
We take that separation as our starting point.
Its phantom-gauge variables are adapted to nonzero charge: the angular eigenvalue and several coefficients diverge as $Q\to0$.
A direct zero-charge substitution therefore does not identify the separated unknowns with the Kerr variables.
For radiative applications, one must also fix the map to physical fields and the normalization at both radial endpoints.
The latter does not follow from a fixed-radius expansion, since coefficients proportional to $Q^2\omega r$ remain important at large radius.

Our aim is to make the separated Kerr--Newman equations usable for physical radiation calculations in a formulation that also includes Kerr.
This requires more than finite coefficients in the separated equations: the master variables must be related to physical curvature and electromagnetic fields, and their boundary normalization must determine the amplitudes of those fields.
We therefore treat equation regularity, field identification, and boundary normalization as parts of the same problem.
The Kerr limit provides a common reference for the gravitational and electromagnetic sectors.

We first organize the separated system in variables that remain finite at zero charge, and then follow these variables through the field transformation and the radial boundary problem.
The physical question is how a separated outgoing mode contributes to gravitational and electromagnetic radiation, including the relative power in the two channels.
Answering it supplies the normalization needed to turn master solutions into radiative observables and prepares the separated formulation for subsequent scattering and response calculations.

\begin{table*}[t]
\caption{Regular first-order systems, final separated equations, physical reconstruction, and boundary-to-radiation dictionary.
Coefficient definitions are given in Eqs.~\eqref{eq:explicit-angular-data}, \eqref{eq:explicit-radial-data}, \eqref{eq:reconstruction-data}, and \eqref{eq:reconstruction-coefficients}; boundary factors are fixed in Eq.~\eqref{eq:boundary-normalization-factors}.
Equation references link to the derivations and applicability conditions.}
\label{tab:results}
\small
\setlength{\tabcolsep}{4pt}
\begin{tabular}{p{.16\textwidth}p{.7\textwidth}p{.085\textwidth}}
\hline\hline\noalign{\vskip4pt}
Result & Formula & Equation \\[4pt]\hline\noalign{\vskip4pt}
Regular first-order systems & \(\displaystyle\begin{aligned}
&\text{Angular:} \qquad \delta_\epsilon\chi_\epsilon=\beta_Q\zeta_\epsilon,\qquad
 \delta_{-\epsilon}\zeta_\epsilon=\gamma_Q\chi_\epsilon,\\
&\text{Radial:} \qquad \mathcal D_\tau W_\tau=F_QV_\tau,\qquad
 \mathcal D_{-\tau}V_\tau=G_QW_\tau,\\
&\sigma=\operatorname{sign}(s),\qquad\epsilon=2s\sigma-3,\qquad\tau=2s-3\sigma.
\end{aligned}\) & \eqref{eq:angular-division-free}, \eqref{eq:regular-radial-pair-unified-D}, \eqref{eq:branch-tau} \\[4pt]\hline\noalign{\vskip4pt}
Angular equation & \(\displaystyle \begin{aligned}
\Biggl[&\frac1{\sin\theta}\partial_\theta(\sin\theta\,\partial_\theta)
 +a^2\omega^2\cos^2\theta-2a\omega s\cos\theta
 +s+A_s-\frac{(m+s\cos\theta)^2}{\sin^2\theta}\\
 &-\frac{2a\omega Q^2}{M^2-Q^2}
   \left(\frac{2a\omega M^2}{M^2-Q^2}-2m+\tau\cos\theta\right)\\
 &+\frac{4\sigma a\omega Q^2}{\mathscr B_{\theta,s}}
   \biggl\{\sin\theta\,\partial_\theta+\tau\biggl[m+s\cos\theta
 -\frac{a\omega(M^2+Q^2)}{M^2-Q^2}\sin^2\theta\biggr]\biggr\}
 \Biggr]S_s=0.
 \end{aligned}\) & \eqref{eq:explicit-teuk-angular} \\[4pt]\hline\noalign{\vskip4pt}
Radial equation & \(\displaystyle \begin{aligned}
 \Biggl[&\Delta_Q^{-s}\partial_r(\Delta_Q^{s+1}\partial_r)
+\frac{K_Q^2-\ii s\Delta_Q'K_Q}{\Delta_Q}
 +4\ii s\omega r-(A_s+a^2\omega^2-2ma\omega)\\
 &-\frac{\omega Q^2}{M^2-Q^2}\left[\frac{4\omega(Mr-Q^2)^2}{M^2-Q^2}
 -\ii(2s+\tau)\Delta_Q'\right]\\
 &-\frac{4\ii\sigma\omega Q^2}{\mathscr B_{r,s}}
  \bigl[\Delta_Q\partial_r+s(1+\tau)(r-M)+\ii\tau K_Q\bigr]
 \Biggr]R_s=0.
 \end{aligned}\) & \eqref{eq:explicit-teuk-radial} \\[4pt]\hline\noalign{\vskip4pt}
Physical fields and Kerr limit & \(\displaystyle\begin{aligned}
\mathcal G_\sigma=\mathcal B_s\Delta_Q^{-2\sigma}g_{\sigma\epsilon},\qquad
\mathcal E_\sigma=\mathcal B_s\Delta_Q^{-\sigma}e_{\sigma\epsilon},\qquad
(\mathcal G_\sigma,\mathcal E_\sigma)|_{Q=0}=\begin{cases}(R_sS_s,0),&|s|=2,\\(0,R_sS_s),&|s|=1.\end{cases}
\end{aligned}\) & \eqref{eq:normalized-radiative-fields}, \eqref{eq:unit-kerr-field} \\[4pt]\hline\noalign{\vskip4pt}
Boundary solutions & \(\displaystyle\begin{aligned}
(R_s^{\mathrm{in}},R_s^{\mathrm{out}})=R_s[\mathsf H]D_\mathrm{H}^{-1},\qquad
(R_s^{\mathrm{down}},R_s^{\mathrm{up}})=R_s[\mathsf I]D_\mathrm{I}^{-1}
\end{aligned}\) & \eqref{eq:normalized-four-solutions} \\[4pt]\hline\noalign{\vskip4pt}
Connection and outgoing amplitude & \(\displaystyle\begin{aligned}
C_T=D_\mathrm{I}C D_\mathrm{H}^{-1},\qquad
A_\mathrm{out}=\frac{1}{(D_\mathrm{I})_{22}}b_\mathrm{up}
\end{aligned}\) & \eqref{eq:kerr-amplitude-dictionary}, \eqref{eq:boundary-to-master-amplitude} \\[4pt]\hline\noalign{\vskip4pt}
Physical outgoing amplitudes & \(\displaystyle\begin{aligned}
 g(\theta)&=\frac{A_{\rm out}}{M^3\sqrt{\sin\theta}}
 \begin{cases}\chi_+,&s=-2,\\Q\zeta_-/(3\sqrt2),&s=-1,\end{cases}\\
 f(\theta)&=-\frac{\ii A_{\rm out}}{M^2\omega\sqrt{\sin\theta}}
 \begin{cases}Q\zeta_+/(3\sqrt2M^2),&s=-2,\\\chi_-/2,&s=-1.
 \end{cases}
\end{aligned}\) & \eqref{eq:physical-outgoing-amplitudes} \\[4pt]\hline\noalign{\vskip4pt}
Outgoing powers & \(\displaystyle\begin{aligned}
 (P_\mathrm G,P_\mathrm E)_{s=-2}&=\frac{|b_{\rm up}|^2}{2M^6\omega^2}
 \left(N_{\chi,+},\frac{Q^2}{9M^2}N_{\zeta,+}\right),\\
 (P_\mathrm G,P_\mathrm E)_{s=-1}&=\frac{|b_{\rm up}|^2}{M^2}
 \left(\frac{Q^2}{9M^2}N_{\zeta,-},N_{\chi,-}\right),\qquad
 N_{v,\epsilon}=\int_0^\pi|v_\epsilon|^2\,\dd\theta
\end{aligned}\) & \eqref{eq:boundary-to-power} \\[4pt]\hline\noalign{\vskip4pt}
Radiation composition & \(\displaystyle
 \left.\frac{P_\mathrm{E}}{P_\mathrm{G}}\right|_{s=-2}
 =\frac{Q^2}{9M^2}\frac{N_{\zeta,+}}{N_{\chi,+}},\qquad 
 \left.\frac{P_\mathrm{G}}{P_\mathrm{E}}\right|_{s=-1}
 =\frac{Q^2}{9M^2}\frac{N_{\zeta,-}}{N_{\chi,-}}.
\) & \eqref{eq:exact-radiation-ratios} \\[4pt]
\hline\hline\noalign{\vskip4pt}
\end{tabular}
\end{table*}

Our scope is the separated radiative subsystem and its physical normalization, rather than full metric reconstruction or a calculation of the scattering spectrum.
The algebraic formulation allows complex frequencies; the angular-continuation and boundary-normalization arguments use the real-frequency assumptions specified in the relevant sections.
The radiation analysis assumes a physical Einstein--Maxwell perturbation with the usual peeling behavior.
For direct use, Table~\ref{tab:results} collects the equations and normalization formulas, with references to their derivations and applicability conditions.

The rest of the paper is organized as follows.
After reviewing the phantom-gauge formulation in Sec.~\ref{sec:review}, in Secs.~\ref{sec:angular} and \ref{sec:radial} we respectively present the separated angular and radial equations that have a smooth Kerr limit.
Section~\ref{sec:physical} identifies the physical fields, and Secs.~\ref{sec:scope} and \ref{sec:flux} establish the boundary and radiation normalizations.
The appendices supply the inverse field transformation, asymptotic flux calculations, and the endpoint-normalization proof.

Throughout the paper, we use geometrized units $c=G=1$ and a Fourier dependence $e^{-\ii\omega t+\ii m\phi}$, with $m\in\mathbb Z$ and possibly complex $\omega$.
Primes on radial and angular functions denote derivatives with respect to their arguments; primes on tetrads or NP components instead label transformed frames, as specified where they occur.

\section{The separated first-order system}\label{sec:review}

We review the Newman--Penrose (NP) formalism~\cite{Newman:1961qr,Chandrasekhar:1985kt} and the separated first-order system of Kerr--Newman perturbations~\cite{Hintz:2026jiw} in the phantom gauge.
Appendix~\ref{app:flux-conventions} collects the NP conventions, and Appendix~\ref{app:physical} gives the separation and inverse field maps.

\subsection{Newman--Penrose formalism}

In Boyer--Lindquist coordinates, with signature $(-+++)$, the background Kerr--Newman metric is
\begin{equation}
 \begin{aligned}
 ds^2={}&-\frac{\Delta_Q}{\Sigma}(dt-a\sin^2\theta\,d\phi)^2
 +\frac{\Sigma}{\Delta_Q}\,dr^2\\
 &+\Sigma\,d\theta^2
 +\frac{\sin^2\theta}{\Sigma}\bigl[(r^2+a^2)d\phi-a\,dt\bigr]^2,
 \end{aligned}
 \label{eq:KN-metric}
\end{equation}
where
\begin{align}
 \Delta_Q&=r^2-2Mr+a^2+Q^2,\\
 \Sigma&=r^2+a^2\cos^2\theta.
\end{align}
The two horizon radii satisfying $\Delta_Q=0$ are
\begin{align}
 r_\pm(Q)=M\pm\sqrt{M^2-a^2-Q^2}.
\end{align}
The background electromagnetic field is $F=dA$, with potential
\begin{equation}
 A_a\dd x^a=-\frac{Qr}{\Sigma}
 (\dd t-a\sin^2\theta\,\dd\phi).
 \label{eq:background-potential}
\end{equation}

The parameters $M>0$, $a$, and $Q$ have dimensions of length, and $\omega$ has dimensions of inverse length.
We write $q=Q^2$, fix $|a|<M$, and take $Q\to0$ within $a^2+Q^2<M^2$.
The limit is therefore uniformly away from extremality for sufficiently small $Q$.
Throughout the paper, we simply write $Q\to0$, but more precisely, we consider $Q/M\to0$.
Charge-order estimates below use the dimensionless parameter $Q/M$, holding $M$, $a/M$, $M\omega$, and, for local radial estimates, $r/M$ fixed.

For later convenience, define
\begin{equation}
 d=r_+-r_-,\qquad x=\frac{r-r_+}{M},\qquad y=\frac{r-r_-}{M}.
 \label{eq:horizon-coordinates}
\end{equation}
For a Fourier mode $e^{-\ii\omega t+\ii m\phi}$, fix the tortoise coordinate and the corotating frequency by
\begin{equation}
\begin{aligned}
 r_*&=r+\frac{r_+^2+a^2}{d}\log x
       -\frac{r_-^2+a^2}{d}\log y,\\
 \Omega_\mathrm{H}&=\frac{a}{r_+^2+a^2},\qquad
 k_\mathrm{H}=\omega-m\Omega_\mathrm{H}.
\end{aligned}
\label{eq:KN-tortoise}
\end{equation}
Then, $\dd r_*/\dd r=(r^2+a^2)/\Delta_Q$ and $r_*=r+2M\log(r/M)+\order{r^{-1}}$.

We use the NP formalism~\cite{Newman:1961qr} and project the perturbations onto the null tetrad $(\ell,n,m,\bar m)$.
For a textbook treatment of the NP formalism and Kerr--Newman geometry, see Ref.~\cite{Chandrasekhar:1985kt}, especially Chaps.~1 and 11.
Here, we state the conventions needed for the separated equations and the later radiative reconstruction.
We use the Kinnersley tetrad
\begin{equation}
 \begin{aligned}
 \ell^a\partial_a&=\frac{r^2+a^2}{\Delta_Q}\partial_t+\partial_r
                   +\frac a{\Delta_Q}\partial_\phi,\\
 n^a\partial_a&=\frac{(r^2+a^2)\partial_t-\Delta_Q\partial_r+a\partial_\phi}{2\Sigma},\\
 m^a\partial_a&=\frac{\ii a\sin\theta\,\partial_t+\partial_\theta+
                         \ii\csc\theta\,\partial_\phi}{\sqrt2\varrho},
 \end{aligned}
 \label{eq:tetrad}
\end{equation}
with
\begin{equation}
\varrho=r+\ii a\cos\theta, \qquad
\varrho^*=r-\ii a\cos\theta.
\label{eq:rho}
\end{equation}

The null tetrad has $\ell\cdot n=-1$ and $m\cdot\bar m=1$.
We adopt the numerical NP components and spin coefficients of Ref.~\cite{Hintz:2026jiw}, translating their defining contractions to our metric signature.
In particular, with the same convention for $R^a{}_{bcd}$, changing $g_{ab}$ to $-g_{ab}$ reverses $C_{abcd}$; the defining signs of the Weyl-scalar contractions are reversed correspondingly, whereas the Maxwell contractions are unchanged.
Thus, the separated equations and the NP field normalizations below are unchanged.
The explicit curvature and spin-coefficient conventions are collected in Appendix~\ref{app:flux-conventions}.

The NP formalism resolves the Weyl and Maxwell tensors into components in this null frame.
Writing $e_1=\ell$, $e_2=n$, $e_3=m$, $e_4=\bar m$, our translated curvature convention gives $\Psi_0=C_{1313}$ and $\Psi_4=C_{2424}$, while $\Phi_0=F_{13}$ and $\Phi_2=F_{42}$.
Their perturbations are the spin-weight $\pm2$ gravitational and spin-weight $\pm1$ electromagnetic radiative components, respectively.
Here, spin weight $s$ means that a component acquires the factor $e^{\ii s\chi}$ under the frame rotation $m\mapsto e^{\ii\chi}m$, with $\ell$ and $n$ fixed.
In the aligned background tetrad, the only nonzero Weyl and Maxwell components are
\begin{equation}
 \Psi_2^{(0)}=\frac{q-M\varrho}{\varrho(\varrho^*)^3},\qquad
 \Phi_1^{(0)}=\frac{Q}{2(\varrho^*)^2}.
 \label{eq:background-NP}
\end{equation}
The superscript $(0)$ denotes a background quantity.
These stationary background components are often called the Coulomb components: their leading large-$r$ terms, $-M/r^3$ and $Q/(2r^2)$, encode the mass and charge fields.
They also determine how an infinitesimal change of null frame mixes the perturbed components.

To specify the frame freedom, let $b_+$ and $b_-$ be infinitesimal complex functions defining null rotations with $n$ and $\ell$ fixed, respectively.
Namely,
\begin{equation}
\begin{aligned}
 n'&=n,\quad m'=m+b_+n,\\
 \ell'&=\ell+\bar b_+m+b_+\bar m,
\end{aligned}
\label{eq:null-rotation-plus}
\end{equation}
and
\begin{equation}
\begin{aligned}
 \ell'&=\ell,\quad m'=m+\bar b_-\ell,\\
 n'&=n+b_-m+\bar b_-\bar m.
\end{aligned}
\label{eq:null-rotation-minus}
\end{equation}
Terms quadratic in the rotation parameters are omitted.
The induced first-order transformations are
\begin{equation}
\begin{aligned}
 \delta\Psi_1'&=\delta\Psi_1+3b_+\Psi_2^{(0)},&
 \delta\Phi_0'&=\delta\Phi_0+2b_+\Phi_1^{(0)},\\
 \delta\Psi_3'&=\delta\Psi_3+3b_-\Psi_2^{(0)},&
 \delta\Phi_2'&=\delta\Phi_2+2b_-\Phi_1^{(0)}.
\end{aligned}
\label{eq:NP-rotation-laws}
\end{equation}
The spin-weight $\pm2$ Weyl perturbations $\delta\Psi_0,\delta\Psi_4$ are unchanged to this order.

For $Q\ne0$, choosing $b_+=-\delta\Phi_0/(2\Phi_1^{(0)})$ and $b_-=-\delta\Phi_2/(2\Phi_1^{(0)})$ imposes the phantom gauge $\delta\Phi_0'=\delta\Phi_2'=0$~\cite{Chandrasekhar:1985kt,Hintz:2026jiw}.
This choice divides by the background charge and cannot define the $Q=0$ variables.
It does not remove electromagnetic radiation: in this gauge that information is carried by the perturbed Weyl components.
In this gauge, following Refs.~\cite{Chandrasekhar:1985kt,Mark:2014aja,Dias:2022oqm}, we use the rescaled Weyl variables
\begin{equation}
\begin{aligned}
 \psi_2&=\delta\Psi_0,&
 \psi_1&=\sqrt2\varrho^*\delta\Psi_1,\\
 \psi_{-2}&=(\varrho^*)^4\delta\Psi_4,&
 \psi_{-1}&=\frac{(\varrho^*)^3}{\sqrt2}\delta\Psi_3.
\end{aligned}
\label{eq:source-NP-variables}
\end{equation}
In particular, $\psi_{\pm1}$ here are rescaled Weyl components in phantom gauge, not Maxwell components in a regular tetrad.

\subsection{Separated first-order system}

Hintz~\cite{Hintz:2026jiw} rewrites the coupled gravitational and electromagnetic perturbation equations in phantom gauge as two first-order systems, one for each sign of the spin weight.
Each system contains two Weyl variables from Eq.~\eqref{eq:source-NP-variables} and two perturbed spin coefficients, which describe the perturbed connection of the null frame.
Linear combinations that diagonalize the radial--angular derivative terms give two two-component fields $y_\pm$.
A matrix integrating factor then gives fields $z_\pm$ whose equations separate into radial and angular parts.
The functions $f_1,f_2$ and $w,w_\sharp$ below are the angular and radial factors of these transformed fields, not of an individual NP scalar.
Appendix~\ref{app:physical} gives the complete transformation and separation ansatz; Sec.~\ref{sec:physical} uses its inverse to recover combinations with finite gravitational and Maxwell limits.
We use the separated systems as the starting point for the zero-charge completion.

To write Hintz's angular system, we introduce the following shorthand for its coefficients:
\begin{equation}
\begin{aligned}
\begin{aligned}
 h&=\frac{3M}{2Q},\\
 t_Q&=\frac32\cot\theta+\sigma m\csc\theta-\sigma c\frac{M^2+Q^2}{M^2-Q^2}\sin\theta,\\
 b_Q&=\frac{2\sigma cM}{M^2-Q^2}\cos\theta,\\
 u_Q&=-h-Qb_Q,\\
 c&=a\omega.
\end{aligned}
\end{aligned}
\label{eq:angular-coefficients}
\end{equation}
The quantities $h,t_Q,u_Q$ are dimensionless, while $b_Q$ has dimensions of inverse length.
The sign $\sigma=\pm1$ distinguishes the positive- and negative-spin systems of Ref.~\cite{Hintz:2026jiw}.
In the Kerr limit it equals $\operatorname{sign}(s)$, where $s=\pm1,\pm2$ is the Teukolsky spin weight; the angular branch introduced in Sec.~\ref{sec:angular} determines its magnitude.

The four-component angular operator and its commuting constant matrix in Steps 2--3 of Ref.~\cite{Hintz:2026jiw} are
\begin{equation}
\begin{aligned}
 \mathcal A_Q&=\begin{pmatrix}\ii\partial_\theta&\ii\Theta_Q\\-\ii\Theta_Q&-\ii\partial_\theta\end{pmatrix},\\
 \Theta_Q&=\begin{pmatrix}t_Q&u_Q\\u_Q&-t_Q\end{pmatrix},\\
 \mathcal S&=\operatorname{diag}(J,-J),\qquad
 J=\begin{pmatrix}0&-1\\1&0\end{pmatrix}.
\end{aligned}
\label{eq:four-angular}
\end{equation}
The derivative entries in $\mathcal A_Q$ multiply the two-dimensional identity.
The matrices obey $[\mathcal A_Q,\mathcal S]=0$ and $\mathcal S^2=-\Id$.
In the basis $(\ii,1,0,0)^T$, $(0,0,-\ii,1)^T$ of $\ker(\mathcal S-\ii)$, the simultaneous angular eigenproblem is
\begin{equation}
 \mathsf A_Q\boldsymbol f=\lambda\boldsymbol f,\qquad
 \mathsf A_Q=
 \begin{pmatrix}
 \ii\partial_\theta&u_Q-\ii t_Q\\
 u_Q+\ii t_Q&-\ii\partial_\theta
 \end{pmatrix}.
 \label{eq:restricted-angular}
\end{equation}
Here, $\boldsymbol f=(f_1,f_2)^T$ contains the two angular component functions, and $\lambda$ is their common separation eigenvalue, to be fixed by the angular endpoint conditions at $\theta=0,\pi$.
Their embedding $\boldsymbol u=(\ii f_1,f_1,-\ii f_2,f_2)^T$ supplies the angular dependence of $z_\pm$ in Eq.~\eqref{eq:transformed-separation}.

The block-swap matrix $\mathcal I=\left(\begin{smallmatrix}0&\Id_2\\\Id_2&0\end{smallmatrix}\right)$ takes an eigenfunction with $(\mathcal S,\lambda)=(\ii,\lambda)$ to one with $(-\ii,-\lambda)$.
The separated radial construction pairs these two angular functions; it does not introduce an additional independent choice of $\lambda$ for the second sector.
Thus, Eq.~\eqref{eq:restricted-angular} retains the first-order constraint that would otherwise be lost by considering only a squared operator.

For the same separation constant $\lambda$, the separated radial pair in Ref.~\cite{Hintz:2026jiw} reads
\begin{equation}
 \begin{aligned}
 w'&=(R_1+\ii\sigma\lambda)w_\sharp,\\
 \Delta_Qw_\sharp'+B_Qw_\sharp&=(R_1-\ii\sigma\lambda)w.
 \end{aligned}
 \label{eq:source-radial}
\end{equation}
The coefficients of this homogeneous radial system are\footnote{The relation between our notation and the quantities $K$ and $\widetilde K_Q$ used in Ref.~\cite{Hintz:2026jiw} is $K_0=-K$ and $K_Q=-\widetilde K_Q$, respectively.}
\begin{equation}
 \begin{aligned}
 R_1&=\ii h-\sigma d_Q,\\
 d_Q&=\frac{2\omega Q(Mr-Q^2)}{M^2-Q^2},\\
 B_Q&=\frac{1-3\sigma}{2}\Delta_Q'-2\ii K_Q,\\
 K_Q&=K_0+\xi_Q\Delta_Q,\\
 K_0&=(r^2+a^2)\omega-am,\\
 \xi_Q&=\frac{2\omega Q^2}{M^2-Q^2}.
 \end{aligned}
 \label{eq:radial-coefficients}
\end{equation}
Primes on radial quantities denote $r$ derivatives; primes on angular quantities below denote $\theta$ derivatives.
The radial factors of $z_\pm$ are related to $w,w_\sharp$ by Eq.~\eqref{eq:radial-factor-map}; its powers of $\Delta_Q$ and phase $H_Q$ account for the normalization of this first-order pair.

Equations~\eqref{eq:restricted-angular} and \eqref{eq:source-radial} are the angular eigenproblem and the radial master system, respectively~\cite{Hintz:2026jiw}; neither contains an inhomogeneous forcing term.
All identities below follow from Eqs.~\eqref{eq:restricted-angular} and \eqref{eq:source-radial}; no small-frequency or slow-rotation approximation is made.
It is explicitly stated in Ref.~\cite{Hintz:2026jiw} that the separated differential system is exact for all complex frequencies, although the mode-stability theorem is restricted to real frequency.
The transformations used here are algebraic differential identities with coefficients analytic in $\omega$, and do not use angular self-adjointness or an energy estimate.
Their validity at complex frequency therefore concerns the equations and changes of variables, not a continuation of the real-frequency spectral or stability conclusions.

\section{Regular angular system and its Kerr limit}\label{sec:angular}

The separated master formulation in Ref.~\cite{Hintz:2026jiw} uses phantom gauge and assumes $Q
\ne0$.
Its coefficients do not permit direct substitution of $Q=0$.
In Secs.~\ref{sec:angular} and \ref{sec:radial}, we extend the phantom-gauge separation to a charge-regular formulation including zero charge by choosing variables and coefficients with finite Kerr limits.

\subsection{A finite spectral parameter}

In the angular system, the parameter $h$ defined in Eq.~\eqref{eq:angular-coefficients} and the coefficients of $\mathsf A_Q$ in Eq.~\eqref{eq:restricted-angular} diverge.
This singular parametrization does not imply the absence of a regular physical Kerr limit,
as Teukolsky's original derivation of the separated equations for $Q=0$ shows~\cite{Teukolsky:1972my,Teukolsky:1973ha}.
Writing $\sigma_j$ for Pauli matrices acting on $\boldsymbol f$, Eq.~\eqref{eq:restricted-angular} gives
\begin{align}
 \mathsf A_Q&=\ii\sigma_3\partial_\theta+t_Q\sigma_2+u_Q\sigma_1.
\end{align}
The divergent part is the constant matrix $-h\sigma_1$, since $u_Q=-h-Qb_Q$.
Its fixed eigenvectors therefore determine the component combinations used below.

The key step to circumvent the singular behavior at $Q\to0$ is to consider the square of $\mathsf A_Q$.
Squaring then isolates its divergent multiple $h^2\Id$ of the identity.
Namely,
\begin{equation}
\begin{aligned}
 \mathsf A_Q^2-h^2\Id
 &=\bigl(-\partial_\theta^2+t_Q^2+3Mb_Q+Q^2b_Q^2\bigr)\Id\\
 &\quad+t_Q'\sigma_1+Qb_Q'\sigma_2.
\end{aligned}
\label{eq:square}
\end{equation}
The cancellation uses $u_Q^2-h^2=3Mb_Q+Q^2b_Q^2$ and $u_Q'=-Qb_Q'$.
Every coefficient in Eq.~\eqref{eq:square} has a finite limit on the open angular interval for $Q\to0$.
The natural finite parameter is therefore
\begin{equation}
 \begin{aligned}
 \mu&=\lambda^2-h^2,\\
 \lambda&=\epsilon hL,\qquad \epsilon=\pm1,\\
 L&=\sqrt{1+\frac{4Q^2\mu}{9M^2}},
 \end{aligned}
 \label{eq:spectral-parameter}
\end{equation}
where the square-root branch is fixed by $L\to1$.
We call the choice $\lambda/h\to\epsilon$ an angular branch; it is distinct from the angular overtone label.
We work locally with bounded $\mu$ and fixed finite $\omega$ as $Q\to0$.
When following a separation-parameter curve, we denote its limiting value by $\mu_0=\mu|_{Q=0}$.
For signed $Q$, $h$ is signed as well: $\epsilon$ labels the ratio $\lambda/h$, not necessarily the sign of $\lambda$.

Equation~\eqref{eq:square} identifies a finite parameter, but an eigenfunction of the square alone need not satisfy the original first-order equation with a chosen $\epsilon$.
We resolve this issue explicitly next.

\subsection{Regular angular system and second-order equation}

To retain the first-order constraint while removing its divergent coefficient, we use the eigenvectors of the constant matrix $-h\sigma_1$ and write
\begin{equation}
 \boldsymbol f=\frac{p}{\sqrt2}\begin{pmatrix}1\\1\end{pmatrix}+\frac{n}{\sqrt2}\begin{pmatrix}1\\-1\end{pmatrix}.
 \label{eq:mass-basis}
\end{equation}
The angular system \eqref{eq:restricted-angular} then becomes
\begin{equation}
 \begin{aligned}
 (u_Q-\lambda)p+\ii(\partial_\theta+t_Q)n&=0,\\
 \ii(\partial_\theta-t_Q)p-(u_Q+\lambda)n&=0.
 \end{aligned}
 \label{eq:pn}
\end{equation}
The same components $p,n$, with these definitions, enter the field inverse in Appendix~\ref{app:physical}.
For a family whose retained component and its first angular derivative remain bounded on a compact interval, $n$ is the leading component for $\epsilon=+1$, while $p$ is smaller by one power of $Q$.
For $\epsilon=-1$, their roles are exchanged; Eq.~\eqref{eq:angular-reconstruction} below makes this scaling precise.

Using $\lambda=\epsilon hL$, both original coefficients factor as
\begin{equation}
\begin{aligned}
 \epsilon=+1:\quad &u_Q-\lambda=-2h\beta_Q,\\
                  &u_Q+\lambda=-\gamma_Q/(2h),\\
 \epsilon=-1:\quad &u_Q+\lambda=-2h\beta_Q,\\
                  &u_Q-\lambda=-\gamma_Q/(2h),
\end{aligned}
\label{eq:angular-branch-factorization}
\end{equation}
where we introduced the angular coupling coefficients
\begin{equation}
 \begin{aligned}
 \beta_Q&=\frac{1+L}{2}+\frac{Q^2b_Q}{3M},\\
 \gamma_Q&=3Mb_Q-\frac{2\mu}{1+L}.
 \end{aligned}
 \label{eq:angular-regular-coefficients}
\end{equation}
Both angular coupling coefficients are dimensionless, with finite $Q\to0$ limits
\begin{equation}
 \beta_Q\to1,\qquad \gamma_Q\to6\sigma c\cos\theta-\mu_0,
 \label{eq:bg-limit}
\end{equation}
and satisfy the useful relation
\begin{equation}
\beta_Q\gamma_Q=-\mu+3Mb_Q+Q^2b_Q^2.
\label{eq:bg-identity}
\end{equation}
Thus, in the first-order system~\eqref{eq:pn}, one coupling diverges as $h$ and the other vanishes as $h^{-1}$.

We absorb the large factor $2h$, together with the phase $\ii$ in Eq.~\eqref{eq:pn}, into the rescaling of the subleading component.
To treat both branches in the same way, retain the leading component and rescale the other component by its vanishing charge factor:
\begin{equation}
 (\chi_\epsilon,\zeta_\epsilon)=
 \begin{cases}(n,-2\ii h p),&\epsilon=+1,\\
 (p,2\ii h n),&\epsilon=-1.
 \end{cases}
 \label{eq:angular-retained-eliminated}
\end{equation}
Thus, the original components are recovered, for $Q\ne0$, as
\begin{equation}
 (n,p)=\begin{cases}
 (\chi_\epsilon,\ii\zeta_\epsilon/(2h)),&\epsilon=+1,\\
 (-\ii\zeta_\epsilon/(2h),\chi_\epsilon),&\epsilon=-1.
 \end{cases}
 \label{eq:angular-source-inverse}
\end{equation}
Introducing the two angular derivative operators
\begin{equation}
 \delta_\pm=\partial_\theta\pm t_Q,
 \label{eq:angular-operators}
\end{equation}
and substituting the rescaled components into Eq.~\eqref{eq:pn} yield
\begin{equation}
 \begin{aligned}
 \delta_\epsilon\chi_\epsilon&=\beta_Q\zeta_\epsilon,\\
 \delta_{-\epsilon}\zeta_\epsilon&=\gamma_Q\chi_\epsilon.
 \end{aligned}
 \label{eq:angular-division-free}
\end{equation}
This is the charge-regular angular first-order system.
It involves no division by either coupling coefficient and defines the regular component pair at $Q=0$ itself.

Note that the rescaling~\eqref{eq:angular-retained-eliminated} is an invertible change of variables only for $Q\ne0$; it is not evaluated as an infinite factor times a vanishing original component at $Q=0$.
At zero charge we instead define $(\chi_\epsilon,\zeta_\epsilon)$ as solutions of the limiting first-order system.
Since $\beta_0=1$, the first equation of \eqref{eq:angular-division-free} gives explicitly
\begin{equation}
 \zeta_\epsilon(0,\theta)
 =\bigl(\partial_\theta+\epsilon t_0\bigr)\chi_\epsilon(0,\theta).
 \label{eq:angular-zero-component}
\end{equation}
Prescribing component data continuous in $Q^2$ at an interior angle yields solutions continuous in $Q^2$ on compact interior intervals.
The regular endpoint eigenmodes require the additional argument in Sec.~\ref{sec:angular-branch}.

On any region where $\beta_Q\ne0$, eliminating $\zeta_\epsilon$ gives the second-order equation and its component reconstruction:
\begin{align}
 &\delta_{-\epsilon}\left[\beta_Q^{-1}\delta_\epsilon\chi_\epsilon\right]
 -\gamma_Q\chi_\epsilon=0,
 \label{eq:angular-regular}\\
 &\zeta_\epsilon=\beta_Q^{-1}\delta_\epsilon\chi_\epsilon.
 \label{eq:angular-reconstruction}
\end{align}
Both relations have a regular zero-charge limit.
Together with Eq.~\eqref{eq:angular-source-inverse}, they recover the original solution at nonzero charge.

These formulas prove equivalence of local solution spaces at nonzero $Q$, rather than only equality of characteristic equations.
Because $\beta_Q\to1$ uniformly in $\theta$ for bounded parameters, its nonvanishing is automatic for sufficiently small $|Q|$.
The coefficients of Eq.~\eqref{eq:angular-regular} are analytic in $Q^2$ near zero, treating $\mu$ and $\omega$ as independent bounded parameters.
In particular, the regularization treats both signs of charge in the same second-order equation.

For later comparison, it is useful to write Eq.~\eqref{eq:angular-regular} as
\begin{equation}
 \begin{split}
& \bigl[-\partial_\theta^2+q_Q\partial_\theta+t_Q^2-\epsilon t_Q'
 +\epsilon q_Qt_Q\\
&\qquad +3Mb_Q+Q^2b_Q^2\bigr]\chi_\epsilon=\mu\chi_\epsilon,
 \end{split}
 \label{eq:angular-scalar}
\end{equation}
where
\begin{equation}
q_Q=\frac{\beta_Q'}{\beta_Q}
= -\frac{2\sigma c q}{3\beta_Q(M^2-q)}\sin\theta.
 \label{eq:qQ}
\end{equation}
The identity~\eqref{eq:bg-identity} is responsible for the simple potential in this expression.
At finite charge, $\beta_Q$ depends on $\mu$ through $L$, so Eq.~\eqref{eq:angular-scalar} is a nonlinear spectral equation.
Its regular appearance does not turn it into a parameter-independent Sturm--Liouville eigenproblem.

\subsection{Angular equation in the Kerr limit}

We now take the Kerr limit $Q\to0$ and denote the zero-charge value of a quantity carrying a $Q$ subscript by a subscript $0$.
For instance,
\begin{equation}
 t_0=t_Q|_{Q=0}=\frac32\cot\theta+\sigma(m\csc\theta-c\sin\theta).
 \label{eq:t0}
\end{equation}
We use $\mu_0$ for the limiting value of $\mu$ defined above.
Taking $Q\to0$ in Eq.~\eqref{eq:angular-scalar} gives
\begin{equation}
 [-\partial_\theta^2+t_0^2-\epsilon t_0'+6\sigma c\cos\theta] \chi_\epsilon=\mu_0\chi_\epsilon.
 \label{eq:angular-limit}
\end{equation}
The standard spin-weighted spheroidal harmonic equation, with the Fourier convention fixed above, is~\cite{Teukolsky:1973ha}
\begin{equation}
 \begin{split}
&\frac1{\sin\theta}\partial_\theta(\sin\theta\,\partial_\theta S_s)\\
 &\quad+\left[c^2\cos^2\theta-2cs\cos\theta+s+A_s\right.\\
 &\qquad\left.-\frac{(m+s\cos\theta)^2}{\sin^2\theta}\right]S_s=0.
 \end{split}
 \label{eq:teuk-angular}
\end{equation}
Here, the separation constant and the spheroidal harmonics are ${}_sA_{\ell m}(c)$ and ${}_sS_{\ell m}(\theta;c)$, respectively.
Throughout the paper, we often omit the labels $\ell,m$ for simplicity.
To compare with Eq.~\eqref{eq:angular-limit}, we first remove the $\cot\theta\,S_s'$ term from Eq.~\eqref{eq:teuk-angular}.
Writing $\chi=\sqrt{\sin\theta}\,S_s$ gives an equation with no first derivative, called the Liouville normal form:
\begin{equation}
 \begin{split}
 \biggl[-\partial_\theta^2+\frac{(m+s\cos\theta)^2-1/4}{\sin^2\theta}
 &-\frac14\\
 {}+2cs\cos\theta-c^2\cos^2\theta-s\biggr]\chi&=A_s\chi.
 \end{split}
 \label{eq:angular-liouville}
\end{equation}

Thus, the required identification is
\begin{equation}
\chi_\epsilon=\sqrt{\sin\theta}\,S_s.
\end{equation}
This factor is fixed by the first-derivative coefficient, up to a constant normalization.
Substituting it into Eq.~\eqref{eq:angular-limit} and comparing its coefficients with Eq.~\eqref{eq:teuk-angular} gives
\begin{equation}
 \begin{array}{c|c|c}
 \epsilon&s&\lambdabar \\ \hline
 +1&2\sigma&\mu_0-2-2\sigma\\
 -1&\sigma&\mu_0+1-\sigma
 \end{array},
 \label{eq:dictionary}
\end{equation}
where
\begin{equation}
\lambdabar \coloneqq A_s+c^2-2mc.
\end{equation}
The angular equation thus identifies the $\epsilon=+1$ branch with the gravitational spin and the $\epsilon=-1$ branch with the electromagnetic spin at the operator level.
The coefficient comparison is an exact operator identity: the constant offset between the two potentials is $2+2\sigma$ for $\epsilon=+$ and $\sigma-1$ for $\epsilon=-$.
Section~\ref{sec:physical} establishes a different statement: which physical NP fields these solutions represent, with their normalization fixed.

The angular dictionary~\eqref{eq:dictionary} fixes both the spin label and the separation-constant shift.
Since $s=\sigma(3+\epsilon)/2$, the inverse branch map is
\begin{equation}
 \sigma=\operatorname{sign}(s),\quad
 \epsilon=2s\sigma-3,\quad
 \tau\coloneqq\sigma\epsilon=2s-3\sigma.
 \label{eq:branch-tau}
\end{equation}
Here, $s\in\{-2,-1,1,2\}$ and $\tau\in\{-1,1\}$.
The separation parameter is then
\begin{equation}
 \mu_0=A_s(c)+c^2-2mc+s(1+\tau).
 \label{eq:full-dictionary-identity}
\end{equation}
Indeed, $s(1+\tau)$ equals $2+2\sigma$ for $\epsilon=+1$ and $\sigma-1$ for $\epsilon=-1$, reproducing the two shifts in the angular dictionary.
These identities require no radial equation; the radial analysis below tests whether the same dictionary applies there.

The regular Kerr angular solutions satisfy
$S_s=\order{\theta^{|m+s|}}$  at $\theta=0$ and
$S_s=\order{(\pi-\theta)^{|m-s|}}$ at $\theta=\pi$.
These conditions specify the physical limiting boundary-value problem, not arbitrary solutions of Eq.~\eqref{eq:angular-limit}.
For $c=0$, Eq.~\eqref{eq:dictionary} gives $\mu_0=\ell(\ell+1)-2$ in both branches, with $\ell\ge\max(|m|,|s|)$.
The value $\mu_0=0$ is therefore allowed for the electromagnetic $\ell=1$ branch when $|m|\le1$.
Our reduction does not divide by $\mu$, and does not exclude that case.

\subsection{Angular equation at finite charge}\label{sec:finite-angular}

The Kerr identification also supplies useful variables at nonzero charge.
We keep $S_s=\chi_\epsilon/\sqrt{\sin\theta}$ and define $A_s(Q)$ and $\lambdabar(Q)=A_s(Q)+c^2-2mc$ by the same algebraic dictionary~\eqref{eq:dictionary}, now replacing $\mu_0$ by the finite-charge parameter $\mu$.
Thus, $s=\sigma(3+\epsilon)/2$ labels the Kerr spin to which the mode continues.
To obtain the finite-charge equation, substitute this rescaling into Eq.~\eqref{eq:angular-scalar} and separate the Kerr terms using
\begin{equation}
 \begin{aligned}
 t_Q-t_0&=-\frac{2\sigma cq}{M^2-q}\sin\theta,\\
 b_Q-b_0&=\frac{2\sigma cq}{M(M^2-q)}\cos\theta.
 \end{aligned}
 \label{eq:tQ0bQ0}
\end{equation}

For direct use of the finite-charge equation, collect the branch identities~\eqref{eq:branch-tau} and \eqref{eq:full-dictionary-identity} with the explicit coefficient definitions (using $\mu$ in place of $\mu_0$):
\begin{equation}
 \begin{aligned}
 \lambdabar&=A_s+a^2\omega^2-2ma\omega,\\
 L_s&=\sqrt{1+\frac{4Q^2}{9M^2}\bigl[\lambdabar+s(1+\tau)\bigr]},\\
 \mathscr B_{\theta,s}&=6(M^2-Q^2)\beta_Q \\
 &=3(M^2-Q^2)(1+L_s)+4\sigma a\omega Q^2\cos\theta.
 \end{aligned}
 \label{eq:explicit-angular-data}
\end{equation}
Here, $s\in\{-2,-1,1,2\}$, $L_s\to1$ fixes the square root, and the separation constant $A_s={}_sA_{\ell m}(a\omega,Q/M)$ is fixed by regularity at both poles; $\ell$ labels the continuation of a Kerr angular mode.
The eigenvalue cannot in general be replaced by an elementary expression in $\ell$, even in Kerr.
Using these identities and Eq.~\eqref{eq:tQ0bQ0} in the rescaled second-order equation, and collecting its derivative and potential terms, gives
\begin{equation}
 \begin{split}
\Biggl[&\frac1{\sin\theta}\partial_\theta(\sin\theta\,\partial_\theta)
 +a^2\omega^2\cos^2\theta-2a\omega s\cos\theta\\
 &+s+A_s-\frac{(m+s\cos\theta)^2}{\sin^2\theta}\\
 &-\frac{2a\omega Q^2}{M^2-Q^2}
   \left(\frac{2a\omega M^2}{M^2-Q^2}-2m+\tau\cos\theta\right)\\
 &+\frac{4\sigma a\omega Q^2}{\mathscr B_{\theta,s}}
   \biggl\{\sin\theta\,\partial_\theta+\tau\biggl[m+s\cos\theta\\
 &\hspace{68pt}-\frac{a\omega(M^2+Q^2)}{M^2-Q^2}\sin^2\theta\biggr]\biggr\}
 \Biggr]S_s=0.
 \end{split}
 \label{eq:explicit-teuk-angular}
\end{equation}

Equation~\eqref{eq:explicit-teuk-angular} is therefore the standard spin-weighted spheroidal operator plus explicit charge corrections, with no singular eigenvalue $\lambda$ left in it.
The charge corrections vanish term by term at $Q=0$.
The corrections do not alter the pole exponents when $\mathscr B_{\theta,s}\ne0$; the regular boundary conditions stated above continue to apply.
At finite charge, $\mathscr B_{\theta,s}$ depends on $A_s$ through $L_s$, making this a nonlinear angular eigenvalue problem.
All explicit charge dependence is through $q=Q^2$.
The next subsection shows that each simple real-frequency Kerr angular eigenvalue also continues as an analytic function of $q$ near zero.

\subsection{Regular angular eigenmodes near zero charge}\label{sec:angular-branch}

The charge-regular equations must describe modes satisfying both angular endpoint conditions at $\theta=0,\pi$, not merely local differential equations.
Fix real $c=a\omega$, $m$, $\sigma$, and $\epsilon$, and a simple Kerr eigenvalue $\mu_0$ in Eq.~\eqref{eq:dictionary}.
Here, simple means multiplicity one at fixed spin and $m$; degeneracies between different $m$ values in the spherical limit are irrelevant.
There is a unique local separation-parameter branch $\mu(q)$, holomorphic near $q=0$, that approaches $\mu_0$ and admits a regular angular eigenfunction at both poles.
For real $q$, this eigenvalue is real.
The reconstructed pair $(\chi_\epsilon,\zeta_\epsilon)$ can be chosen holomorphic in $L^2(0,\pi;\dd\theta)$ and converges to the Kerr pair.
For nonzero real $Q$, its inverse map belongs to the original first-order angular domain and has eigenvalue $\lambda=\epsilon hL(\mu(q))$.

Appendix~\ref{app:angular-proof} proves this local continuation of the regular angular eigenvalue and eigenfunction, including the first-order endpoint conditions and normalization freedom.
The proof keeps the singular pole terms fixed and treats the charge-dependent potential as a bounded perturbation.
Because that potential also depends on the unknown separation parameter, an implicit-function argument is needed to impose spectral self-consistency.
The inverse first-order map requires a further endpoint check: the potentially nonintegrable term cancels at the critical poles $m\pm s=0$.

This result supplies an actual separation curve for the boundary analysis in Sec.~\ref{sec:scope}, and angular functions with the norms required by the radiation formulas in Sec.~\ref{sec:flux}.
It concerns each fixed mode near zero charge; it does not assert uniform continuation of the entire spectrum, global analyticity in complex frequency, or full spacetime reconstruction.

\section{Regular radial system and its Kerr limit}\label{sec:radial}

\subsection{Branch-dependent rescaling at finite charge}

Let us consider the Kerr limit $Q\to0$ of the radial equation \eqref{eq:source-radial}.
The branch $\epsilon$ determines which coupling coefficient diverges:
\begin{equation}
R_1\pm\ii\sigma\lambda=\ii h(1\pm\sigma\epsilon)+\order{Q/M},
\label{eq:radial-rhs-limit}
\end{equation}
at fixed $r$.
One coefficient vanishes as $Q/M$, while the other grows as $M/Q$.
Thus, the original component normalization does not admit direct substitution of $Q=0$.

This observation identifies the large coefficient and therefore determines which radial component must be rescaled.
The product $\sigma\epsilon$ in Eq.~\eqref{eq:radial-rhs-limit} is precisely the sign $\tau=2s-3\sigma$ introduced in Eq.~\eqref{eq:branch-tau}.
We therefore classify the radial rescaling by the same $\tau=\pm1$.

Substituting $\lambda=\epsilon hL$ and $R_1=\ii h-\sigma d_Q$ and using $L-1=4Q^2\mu/[9M^2(1+L)]$, we factor the coefficients exactly as
\begin{align}
 \tau&=+1:\quad\notag\\
 &R_1+\ii\sigma\lambda = 2\ii hF_Q, \qquad R_1-\ii\sigma\lambda=G_Q/(2\ii h),
 \label{eq:rhs-plus}\\
 \tau&=-1:\quad\notag\\
 &R_1+\ii\sigma\lambda = G_Q/(2\ii h), \qquad R_1-\ii\sigma\lambda=2\ii hF_Q,
 \label{eq:rhs-minus}
\end{align}
where we introduced the radial coupling coefficients
\begin{equation}
\begin{aligned}
 F_Q=\frac{1+L}{2}+\frac{\ii\sigma d_Q}{2h},\qquad
 G_Q=\frac{2\mu}{1+L}-2\ii\sigma h d_Q.
\end{aligned}
\label{eq:FG}
\end{equation}
Both functions are dimensionless and analytic in $Q^2$ at fixed $r$.
In particular, at fixed $r$ as $Q\to0$,
\begin{equation}
 F_Q\to1,\qquad
 G_Q\to P(r)\coloneqq \mu_0-6\ii\sigma\omega r.
 \label{eq:FG-limit}
\end{equation}
Note that they satisfy a useful relation
\begin{equation}
 F_QG_Q=\mu+d_Q^2-2\ii\sigma h d_Q.
 \label{eq:FG-identity}
\end{equation}

These finite coefficients suggest absorbing the remaining powers of $2\ii h$ into the radial component.
We thus define a new variable
\begin{equation}
 w_\flat=\begin{cases}
 2\ii hMw_\sharp,&\tau=+1,\\
 Mw_\sharp/(2\ii h),&\tau=-1.
 \end{cases}
 \label{eq:radial-rescaling}
\end{equation}
Here, we multiply by $M$ in Eq.~\eqref{eq:radial-rescaling} so that $w_\flat$ has the same dimension as $w$.
The original $w_\sharp$ has dimensions of $w$ divided by length, as follows from Eq.~\eqref{eq:source-radial}.
Similarly, we define the two dimensionless derivative operators appearing in the first-order system:
\begin{equation}
 \mathcal D_+=M\partial_r,\qquad
 \mathcal D_-=M^{-1}(\Delta_Q\partial_r+B_Q).
 \label{eq:radial-operators}
\end{equation}
Then, the radial first-order system~\eqref{eq:source-radial} becomes, exactly,
\begin{equation}
 \begin{aligned}
\tau=+1:\quad&\begin{cases}\mathcal D_+ w=F_Qw_\flat,\\ \mathcal D_- w_\flat=G_Qw,\end{cases}\\[3pt] \tau=-1:\quad&\begin{cases}\mathcal D_+ w=G_Qw_\flat,\\\mathcal D_- w_\flat=F_Qw.\end{cases}
 \end{aligned}
 \label{eq:regular-radial-pair}
\end{equation}

We can further rewrite the system~\eqref{eq:regular-radial-pair} in the following two unified forms.
First, by defining
\begin{equation}
 (\mathcal P_\tau,\mathcal R_\tau)=
 \begin{cases}(F_Q,G_Q),&\tau=+1,\\(G_Q,F_Q),&\tau=-1,\end{cases}
 \label{eq:radial-coupling-coefficients-renamed}
\end{equation}
we can rewrite Eq.~\eqref{eq:regular-radial-pair} as
\begin{equation}
 \begin{aligned}
 \mathcal D_+ w&=\mathcal P_\tau w_\flat,\\
 \mathcal D_- w_\flat&=\mathcal R_\tau w.
 \end{aligned}
 \label{eq:regular-radial-pair-unified}
\end{equation}
Second, by defining the retained and eliminated components by
\begin{equation}
 (W_\tau,V_\tau)=
 \begin{cases}(w,w_\flat),&\tau=+1,\\(w_\flat,w),&\tau=-1,\end{cases}
 \label{eq:radial-retained-eliminated}
\end{equation}
Eq.~\eqref{eq:regular-radial-pair} reads
\begin{equation}
 \begin{aligned}
\mathcal D_\tau W_\tau&=F_QV_\tau,\\
\mathcal D_{-\tau}V_\tau&=G_QW_\tau.
 \end{aligned}
 \label{eq:regular-radial-pair-unified-D}
\end{equation}

Equations~\eqref{eq:regular-radial-pair}, \eqref{eq:regular-radial-pair-unified}, and \eqref{eq:regular-radial-pair-unified-D} are three equivalent representations of the same charge-regular radial master system.
The branchwise form~\eqref{eq:regular-radial-pair} displays the component rescaling directly.
The fixed-order form~\eqref{eq:regular-radial-pair-unified} keeps $Y=(w,w_\flat)^T$ in the same order for both branches; we use it in Sec.~\ref{sec:scope} to construct the horizon and infinity bases and their connection matrix.
The retained-component form~\eqref{eq:regular-radial-pair-unified-D} keeps the couplings in the order $(F_Q,G_Q)$ and makes the correspondence with the angular system~\eqref{eq:angular-division-free} explicit:
\begin{equation}
\begin{aligned}
(\chi_\epsilon,\zeta_\epsilon)&\longleftrightarrow(W_\tau,V_\tau),\\
(\beta_Q,\gamma_Q)&\longleftrightarrow(F_Q,G_Q),\\
(\delta_\epsilon,\delta_{-\epsilon})&\longleftrightarrow(\mathcal D_\tau,\mathcal D_{-\tau}).
\end{aligned}
\label{eq:first-order-correspondence}
\end{equation}
We therefore use Eq.~\eqref{eq:regular-radial-pair-unified-D} as the representative radial system in Table~\ref{tab:results} and for the second-order reduction below.

The angular system~\eqref{eq:angular-division-free} and any of these three radial forms provide the separated formulation including the Kerr limit.
The coefficients of the angular and radial first-order systems remain finite as $Q\to0$, with $r$ fixed in the exterior.
The rescaling concerns the relative normalization of components, not a boundary condition.
Consequently, these systems describe arbitrary local homogeneous solutions, including general frequencies away from quasinormal modes (QNMs).

As in the angular construction, the rescaling~\eqref{eq:radial-rescaling} is used only for $Q\ne0$.
At $Q=0$, $(w,w_\flat)$ are defined by the limiting system~\eqref{eq:regular-radial-pair}, with finite initial data at an exterior reference radius.
For $\tau=+1$ its first equation gives $w_\flat=Mw'$; for $\tau=-1$ its second equation gives $w=M^{-1}(\Delta_0w_\flat'+B_0w_\flat)$.
Neither definition requires dividing by the possibly vanishing $G_0$.
Continuous initial data give a continuous local family in $Q^2$; the separate question of normalization at infinity is addressed in Sec.~\ref{sec:scope}.

To obtain a single second-order equation, we eliminate through $F_Q$, which tends to one, rather than through $G_Q$, which can vanish in the Kerr limit.
For this elimination, we use the retained-component form~\eqref{eq:regular-radial-pair-unified-D} that isolates $F_Q$ and $G_Q$ explicitly.
For either $\tau=\pm1$, the first equation reconstructs the second component and the second equation gives the second-order reduction.

Solving the first relation and substituting into the second gives the unified radial equation and inverse map:
\begin{align}
 &\mathcal D_{-\tau}\left[F_Q^{-1}\mathcal D_\tau W_\tau\right]
 -G_QW_\tau=0,\label{eq:radial-unified}\\
 &V_\tau=F_Q^{-1}\mathcal D_\tau W_\tau.
 \label{eq:radial-inverse}
\end{align}
Their structure parallels Eqs.~\eqref{eq:angular-regular} and \eqref{eq:angular-reconstruction}.
They are equivalent to the first-order system wherever $F_Q\ne0$, and remain nonsingular near zero charge on each fixed compact exterior interval, including when $\omega=0$ or $\mu=0$.

For coefficient comparisons, put $\bar\tau=(1-\tau)/2$  and suppress the branch subscript on $W$.
Expanding Eq.~\eqref{eq:radial-unified} gives
\begin{equation}
 \begin{split}
 \Delta_QW''+\left(B_Q+\bar\tau\Delta_Q'-\Delta_Q\frac{F_Q'}{F_Q}\right)W'&\\
 +\left[\bar\tau\left(B_Q'-B_Q\frac{F_Q'}{F_Q}\right)-F_QG_Q\right]W&=0.
 \end{split}
 \label{eq:radial-unified-expanded}
\end{equation}
The branch-dependent derivative terms arise from the order of $\mathcal D_+$ and $\mathcal D_-$ in the factorized equation.

\subsection{Four-spin correspondence in the Kerr limit}

Let us check the Kerr limit $Q\to0$.
Let $\Delta_0=r^2-2Mr+a^2$ and let $(s,\lambdabar)$ be given by Eq.~\eqref{eq:dictionary}.
Equation~\eqref{eq:radial-unified-expanded} reduces in all four cases to the same horizon-factored expression,
\begin{equation}
 \begin{split}
 \Delta_0W''+[(1-s)\Delta_0'-2\ii K_0]W'&\\
 {}+[\ii(4s-2)\omega r-\lambdabar-2s]W&=0.
 \end{split}
 \label{eq:radial-limit}
\end{equation}
For transparency, the full dictionary is
\begin{equation}
 \begin{array}{cc|c|c|c}
 \sigma&\epsilon&s&\lambdabar&W\\ \hline
 +1&+1&+2&\mu_0-4&w\\
 +1&-1&+1&\mu_0&w_\flat\\
 -1&+1&-2&\mu_0&w_\flat\\
 -1&-1&-1&\mu_0+2&w
 \end{array}.
 \label{eq:full-dictionary}
\end{equation}
The radial dictionary~\eqref{eq:full-dictionary} agrees with Eq.~\eqref{eq:full-dictionary-identity}, already derived from the angular equation.
The angular and radial equations therefore select the same Kerr separation constant in every branch; the match is not achieved by choosing independent angular and radial shifts.

The radial factor is likewise fixed by matching the first-derivative coefficient.
Writing $R_s=f_sW$, comparison with the Teukolsky coefficient $(s+1)\Delta_0'$ requires
\begin{equation}
 \frac{f_s'}{f_s}=-s\frac{\Delta_0'}{\Delta_0}-\ii\frac{K_0}{\Delta_0}.
\end{equation}
Its integral gives the factor below; the remaining zeroth-order coefficient provides a separate consistency check.

Integrating this logarithmic derivative gives
\begin{equation}
 R_s\propto\Delta_0^{-s}\exp\left[-\ii\int^r
 \frac{K_0(r')}{\Delta_0(r')}\,\dd r'\right]W.
\end{equation}
Thus, the transformation is fixed by coefficient matching up to an $r$-independent factor.
To fix the integration constant as well, we choose the phase with unit asymptotic coefficient:
\begin{align}
 H_0(r)&=e^{-\ii\omega r}(r/M)^{-2\ii M\omega}\notag\\
 &\quad\times\exp\left[\ii\int_r^\infty
 \left(\frac{K_0(r')}{\Delta_0(r')}-\omega-\frac{2M\omega}{r'}\right)\dd r'\right], \notag\\
 R_s&=H_0(\Delta_0/M^2)^{-s}W.
 \label{eq:radial-conjugation}
\end{align}
The subtracted integrand is $\order{r^{\prime-2}}$.
The factor $\Delta_0/M^2$ is dimensionless, so $R_s$ and $W$ have the same dimensions.
This fixes the dependent-variable map, while boundary amplitudes remain freely specifiable.
Substituting Eq.~\eqref{eq:radial-conjugation} into Eq.~\eqref{eq:radial-limit} gives
\begin{equation}
 \begin{split}
 &\Delta_0^{-s}\partial_r(\Delta_0^{s+1}\partial_rR_s)\\
 &\quad+\left[\frac{K_0^2-\ii s\Delta_0'K_0}{\Delta_0}
 +4\ii s\omega r-\lambdabar\right]R_s=0.
 \end{split}
 \label{eq:teuk-radial}
\end{equation}
which is the radial Teukolsky equation for Kerr in the convention of Eq.~\eqref{eq:teuk-angular}~\cite{Teukolsky:1973ha}.
The zeroth-order coefficient after conjugation is $-s\Delta_0''-\ii K_0'+4\ii s\omega r-\lambdabar$; using $\Delta_0''=2$ and $K_0'=2\omega r$ gives Eq.~\eqref{eq:radial-limit}, including its constant shift $-2s$.

This fixes the radial variable, including its constant phase and mass normalization.
Section~\ref{sec:physical} identifies its NP field normalization, and Sec.~\ref{sec:kerr-amplitudes} fixes its horizon and infinity amplitudes explicitly.
The finite-charge outgoing flux is then obtained in Sec.~\ref{sec:flux}.

\subsection{Radial equation at finite charge}\label{sec:finite-radial}

We now extend the radial variable itself to finite charge, so that both separated equations display their Kerr limits directly.
The first-derivative coefficient in Eq.~\eqref{eq:radial-unified-expanded} contains $K_Q$, suggesting the same conjugation as at zero charge, \eqref{eq:radial-conjugation}, with $K_0$ replaced by $K_Q$.
The same coefficient matching gives
\begin{equation}
 R_s\propto\Delta_Q^{-s}\exp\left[-\ii\int^r
 \frac{K_Q(r')}{\Delta_Q(r')}\,\dd r'\right]W_\tau.
\end{equation}
We fix the remaining constant by the dimensionless radial convention and unit asymptotic phase used below.
Using $d,x,y$ from Eq.~\eqref{eq:horizon-coordinates}, define
\begin{equation}
\begin{aligned}
 H_Q(r)&=e^{-\ii(\omega+\xi_Q)r}x^{-\ii\alpha_+}y^{-\ii\alpha_-},\\
 \alpha_+&=\frac{K_0(r_+)}d,\qquad
 \alpha_-=-\frac{K_0(r_-)}d,\\
 R_s(Q,r)&=H_Q(r)(\Delta_Q/M^2)^{-s}W_\tau(r).
\end{aligned}
\label{eq:finite-teuk-variable}
\end{equation}
Here, $x,y>0$ in the exterior, and their real logarithms specify the powers.
Since $H_Q'/H_Q=-\ii K_Q/\Delta_Q$ and $\alpha_++\alpha_-=2M\omega$, this phase agrees with Eq.~\eqref{eq:radial-conjugation} at $Q=0$ and obeys
\begin{equation}
 \lim_{r\to\infty}H_Q(r)e^{\ii(\omega+\xi_Q)r}
 (r/M)^{2\ii M\omega}=1.
 \label{eq:phase-at-infinity}
\end{equation}

Substituting $W_\tau=H_Q^{-1}(\Delta_Q/M^2)^s R_s$ into Eq.~\eqref{eq:radial-unified-expanded},
the logarithmic derivative of its prefactor, $s\Delta_Q'/\Delta_Q+\ii K_Q/\Delta_Q$, fixes the derivative terms.
To simplify the potential, we use Eqs.~\eqref{eq:radial-coefficients}, \eqref{eq:full-dictionary-identity}, \eqref{eq:FG-identity}, and $F_Q'=2\ii\sigma\omega q/[3(M^2-q)]$.
For an explicit expression, take $\sigma,\tau,L_s,\lambdabar$ from Eq.~\eqref{eq:explicit-angular-data} and set
\begin{equation}
 \begin{aligned}
 \Delta_Q&=r^2-2Mr+a^2+Q^2,\\
 K_Q&=(r^2+a^2)\omega-am+\frac{2\omega Q^2}{M^2-Q^2}\Delta_Q,\\
 \mathscr B_{r,s}&=6(M^2-Q^2)F_Q \\
 &=3(M^2-Q^2)(1+L_s)+4\ii\sigma\omega Q^2\left(r-\frac{Q^2}{M}\right).
 \end{aligned}
 \label{eq:explicit-radial-data}
\end{equation}
The remaining logarithmic derivative can also be eliminated, giving the exact radial equation
\begin{equation}
 \begin{split}
 \Biggl[&\Delta_Q^{-s}\partial_r(\Delta_Q^{s+1}\partial_r)
 +\frac{K_Q^2-\ii s\Delta_Q'K_Q}{\Delta_Q}
 +4\ii s\omega r-\lambdabar \\
 &-\frac{\omega Q^2}{M^2-Q^2}\left( \frac{4\omega(Mr-Q^2)^2}{M^2-Q^2}
 -\ii(2s+\tau)\Delta_Q' \right)\\
 &-\frac{4\ii\sigma\omega Q^2}{\mathscr B_{r,s}}
  \bigl[\Delta_Q\partial_r+s(1+\tau)(r-M)+\ii\tau K_Q\bigr]
 \Biggr]R_s=0.
 \end{split}
 \label{eq:explicit-teuk-radial}
\end{equation}

The substitutions above give this equation directly from the regular radial system on regions where $F_Q\ne0$.
Every additional term carries $Q^2$, and $K_Q\to(r^2+a^2)\omega-am$ at zero charge, so direct substitution gives Eq.~\eqref{eq:teuk-radial}.

Together with the angular equation~\eqref{eq:explicit-teuk-angular}, the radial equation~\eqref{eq:explicit-teuk-radial} depends explicitly on the charge only through $Q^2$.
This reflects charge reversal in the source-free Einstein--Maxwell system: reversing the background charge and the Maxwell perturbation leaves the metric perturbation and radiative boundary conditions unchanged.
Consequently, the QNM spectrum as a set is invariant under $Q\to-Q$, consistently with the weak-charge formulation of Ref.~\cite{Mark:2014aja}.
This symmetry alone does not prove analyticity of individual QNM branches at degeneracies; the local angular result in Sec.~\ref{sec:angular-branch} does not assume such a QNM continuation.

A useful normalization invariant follows without solving the radial equation.
For two solutions $R_{s,1}$ and $R_{s,2}$ at the same $(\omega,\mu,\sigma,\epsilon)$, Abel's identity gives
\begin{equation}
 \mathcal W_s=\frac{\Delta_Q^{s+1}}{F_Q}
 (R_{s,1}R_{s,2}'-R_{s,2}R_{s,1}'),\qquad
 \mathcal W_s'=0.
 \label{eq:finite-teuk-wronskian}
\end{equation}
It reduces to the Teukolsky Wronskian at zero charge.
This bilinear invariant checks connection coefficients; it is not by itself an energy-flux conservation law. Section~\ref{sec:flux} derives the physical outgoing fluxes; a horizon-inclusive energy balance is beyond the present analysis.

\subsection{Comparison with a spin-zero test field}
\label{sec:spin-zero}

The derivation in Secs.~\ref{sec:angular} and \ref{sec:radial} concerns the coupled Einstein--Maxwell radiative sectors $s=\pm1,\pm2$.
A neutral, massless test field satisfying $\Box_g\Phi=0$ on Kerr--Newman is separately known to admit separation~\cite{Carter:1968ks,Boonserm:2014rma}.
For $\Phi=e^{-\ii\omega t+\ii m\phi}R_0(r)S_0(\theta)$, its equations in our conventions are
\begin{align}
 &\biggl[\frac1{\sin\theta}\partial_\theta(\sin\theta\partial_\theta)
 +a^2\omega^2\cos^2\theta\notag\\
 &\hspace{42pt}+A_0-\frac{m^2}{\sin^2\theta}\biggr]S_0=0,
 \label{eq:neutral-scalar-angular}\\
 &\left[\partial_r(\Delta_Q\partial_r)+\frac{K_0^2}{\Delta_Q}
 -\lambdabar_0\right]R_0=0.
 \label{eq:neutral-scalar-radial}
\end{align}
Here, $\lambdabar_0=A_0+a^2\omega^2-2ma\omega$, and $A_0$ is the ordinary spin-zero spheroidal eigenvalue, independent of $Q$ at fixed $a\omega$.

Although $s=0$ is outside our branch dictionary, the explicit operators admit an algebraic check by assigning $s=\sigma=\tau=0$.
Define the coordinate-independent shift
\begin{equation}
 C_Q=\frac{4a\omega Q^2}{M^2-Q^2}
       \left(\frac{a\omega M^2}{M^2-Q^2}-m\right).
 \label{eq:spin-zero-shift}
\end{equation}
The angular correction then equals $-C_Q$, while the radial correction obeys
\begin{equation}
 \frac{K_Q^2-K_0^2}{\Delta_Q}
 -\frac{4\omega^2Q^2(Mr-Q^2)^2}{(M^2-Q^2)^2}=C_Q.
 \label{eq:spin-zero-cancellation}
\end{equation}

Consequently, Eqs.~\eqref{eq:explicit-teuk-angular} and \eqref{eq:explicit-teuk-radial} reproduce the test-field equations above after identifying $A_0=A_s|_{s=\sigma=\tau=0}-C_Q$ or equivalently $\lambdabar_0=\lambdabar|_{s=\sigma=\tau=0}-C_Q$.
Keeping the same numerical separation constant without this shift would not give the conventional spin-zero equations at generic finite $a\omega Q$.
This is a consistency check of the final second-order operators, not an additional branch of the Einstein--Maxwell first-order system or its radiative reconstruction.

\subsection{Summary of the charge-regular separated equations}

Let us summarize the angular and radial equations which we derived in Secs.~\ref{sec:angular} and \ref{sec:radial}.
Equations~\eqref{eq:angular-division-free} and \eqref{eq:regular-radial-pair-unified-D}, with $L(0)=1$, provide the charge-regular separated formulation near $Q=0$.
In the real-frequency neighborhood used for our analyticity results, $L$ is real, $\operatorname{Re}F_Q=(1+L)/2>0$, and $\beta_Q$ stays uniformly away from zero.
Elimination is therefore valid there; retaining the first-order systems also covers parameter domains, including complex frequency, where an elimination coefficient can vanish.
The second-order angular equation~\eqref{eq:angular-regular} divides by $\beta_Q$, and the second-order radial reduction \eqref{eq:radial-unified} divides by $F_Q$.
The first-order systems~\eqref{eq:angular-division-free} and \eqref{eq:regular-radial-pair-unified-D} retain both components without these divisions.
They therefore supply the continuation through zeros of these coefficients, where the second-order reductions and their divided reconstruction formulas cannot be used directly.

There is no need to retain an infinite $\lambda$ as a variable of this completed formulation: its finite coordinates are $\mu$, $\epsilon$, and the regularized unknowns.
For $Q\ne0$ they are exactly related to the first-order system in Ref.~\cite{Hintz:2026jiw}; at $Q=0$ the transformations
\begin{equation}
 S_s=\frac{\chi_\epsilon}{\sqrt{\sin\theta}},\qquad
 R_s=H_0(\Delta_0/M^2)^{-s}W
 \label{eq:joint-teukolsky-map}
\end{equation}
give precisely the angular and radial Teukolsky equations~\eqref{eq:teuk-angular} and \eqref{eq:teuk-radial}, with the common separation constant in Eq.~\eqref{eq:full-dictionary}.

Equations~\eqref{eq:explicit-teuk-angular} and \eqref{eq:explicit-teuk-radial} are the explicit finite-charge Teukolsky form of the separated master equations, with the common dictionary~\eqref{eq:full-dictionary}.
They express the second-order operators directly in physical parameters and one angular eigenvalue, and recover the Kerr equations by setting $Q=0$.
Their angular and radial reductions require $\beta_Q\ne0$ and $F_Q\ne0$ respectively; Eqs.~\eqref{eq:angular-division-free} and \eqref{eq:regular-radial-pair-unified-D} supply the corresponding first-order formulations across zeros of these coefficients.
Indeed, $\mathscr B_{\theta,s}=6(M^2-Q^2)\beta_Q$ and $\mathscr B_{r,s}=6(M^2-Q^2)F_Q$, so these are exactly the nonvanishing-denominator conditions in the explicit equations.

The first-order systems~\eqref{eq:angular-division-free} and \eqref{eq:regular-radial-pair-unified-D}, the explicit second-order equations~\eqref{eq:explicit-teuk-angular} and \eqref{eq:explicit-teuk-radial}, and their Kerr reduction are valid for complex $\omega$ as differential identities.
No step in their derivation requires real $\omega$; the coefficients are analytic in $\omega$ with the separation parameter held independent.
For the spectral square root, we use the branch $L_s\to1$ as $Q\to0$; the second-order forms additionally require the nonvanishing denominators stated above.
This statement does not assume a globally analytic angular eigenvalue branch or completeness of separated modes at complex frequency.
It is distinct from the local angular-branch theorem in Sec.~\ref{sec:angular-branch} and the boundary-normalized connection theorem in Sec.~\ref{sec:scope}, which are proved at fixed real frequency (nonzero for the boundary theorem).
The outgoing energy-flux formulas in Sec.~\ref{sec:flux} likewise concern nonzero real frequency.

The results obtained above establish an exact differential-equation correspondence for all four spins, not a small-frequency or slow-rotation approximation.
The singular inverse transformation to phantom-gauge variables is not used to define the zero-charge system.
The variables coincide with the Teukolsky variables at zero charge, but describe mixed gravitational--electromagnetic modes at nonzero charge; their physical fields are reconstructed in Sec.~\ref{sec:physical}.
No matter-source projection is inferred from the homogeneous equations.

\section{Physical radiative fields and normalization}\label{sec:physical}

We now identify the radiative fields represented by the regular master solutions.
The task has two steps: first construct field combinations unaffected by the null rotations used to impose phantom gauge, then invert the separated transformation and fix their Kerr normalization.
We use the tetrad and background fields of Sec.~\ref{sec:review} and suppress the common Fourier factor.
Appendix~\ref{app:physical} specifies the transformation to the separated variables and its inverse.

\subsection{Construction of radiative fields}

Equation~\eqref{eq:NP-rotation-laws} shows that $Q\delta\Psi_1=2(\varrho^*)^2\Phi_1^{(0)}\delta\Psi_1$ and $3(\varrho^*)^2\Psi_2^{(0)}\delta\Phi_0$ acquire the same shift.
Their difference is therefore invariant under this linearized null rotation; the same cancellation holds for $(\delta\Psi_3,\delta\Phi_2)$.
We choose the overall factors so that the electromagnetic combinations reduce directly to the Kerr Maxwell variables.
This gives the regular radiative combinations
\begin{equation}
\begin{aligned}
 \mathcal G_+&=\delta\Psi_0,\qquad
 \mathcal G_-=(\varrho^*)^4\delta\Psi_4,\\
 \mathcal E_+&=\frac{\varrho^*}{3M}\bigl[Q\delta\Psi_1
 -3(\varrho^*)^2\Psi_2^{(0)}\delta\Phi_0\bigr],\\
 \mathcal E_-&=\frac{(\varrho^*)^3}{3M}\bigl[Q\delta\Psi_3
 -3(\varrho^*)^2\Psi_2^{(0)}\delta\Phi_2\bigr].
\end{aligned}
\label{eq:regular-radiation}
\end{equation}
They retain their spin and boost weights.
A tetrad boost is $\ell^a\mapsto A\ell^a$, $n^a\mapsto A^{-1}n^a$ with $A>0$ and $m^a$ unchanged; a spin rotation is $m^a\mapsto e^{\ii\psi}m^a$.
Invariance here refers only to the infinitesimal null rotations used to impose phantom gauge.

The mixed invariant appears in the Kerr--Newman perturbation treatment reviewed in Ref.~\cite{Chandrasekhar:1985kt} and in the explicitly gauge-invariant formulation of Ref.~\cite{Dias:2022oqm}.
In our conventions, its negative-spin combination is $\varphi_{-1}=2\Phi_1^{(0)}\delta\Psi_3-3\Psi_2^{(0)}\delta\Phi_2$, and $\mathcal E_-=(\varrho^*)^5\varphi_{-1}/(3M)$.

For $Q\ne0$ in that gauge, the phantom-gauge variables defined in Eq.~\eqref{eq:source-NP-variables} obey
\begin{equation}
 \begin{aligned}
 \mathcal G_\pm =\psi_{\pm2},\quad\,
 \mathcal E_+ =\frac{Q}{3\sqrt2M}\psi_1,\quad\,
 \mathcal E_- =\frac{\sqrt2Q}{3M}\psi_{-1}.
\end{aligned}
\label{eq:phantom-radiative-fields}
\end{equation}
In a nonsingular tetrad family the zero-charge values are instead
\begin{equation}
\begin{aligned}
 \mathcal E_+(0)&=\delta\Phi_0,\\
 \mathcal E_-(0)&=(\varrho^*)^2\delta\Phi_2.
\end{aligned}
\label{eq:EM-limit}
\end{equation}
Thus, multiplying the phantom-gauge Weyl variables by $Q$ preserves, rather than discards, the Maxwell radiation in the limit.
Recovering $\psi_{\pm1}$ from these relations requires division by $Q$, so phantom-gauge variables generally diverge for a family with a nonzero Maxwell limit, whereas $\mathcal E_\pm$ remains finite.
At finite charge, $\mathcal E_\pm$ are mixed curvature--Maxwell invariants, not the Maxwell scalars alone.

\subsection{Normalized reconstruction and Kerr correspondence}

We normalize each separated mode so that its nonvanishing radiative field tends to $R_sS_s$ at zero charge.
For the angular pair $(\chi_\epsilon,\zeta_\epsilon)$ and radial pair $(w,w_\flat)$ of Secs.~\ref{sec:angular} and \ref{sec:radial}, define
\begin{equation}
\begin{aligned}
 \mathcal B_s&=\frac{M^{2s}H_Q}{(M^2-q)\sqrt{\sin\theta}},\\
 C_\varrho&=M-q/\varrho,\qquad A_\theta=a\sin\theta.
\end{aligned}
\label{eq:reconstruction-data}
\end{equation}
The factor $M^{2s}$ matches the radial convention in Eq.~\eqref{eq:finite-teuk-variable}.
We give the resulting reconstruction directly; the normalization of its inverse map is specified in Appendix~\ref{app:physical}.
The final reconstruction is then
\begin{equation}
\begin{aligned}
 \mathcal G_\sigma&=\mathcal B_s\Delta_Q^{-2\sigma}g_{\sigma\epsilon},\\
 \mathcal E_\sigma&=\mathcal B_s\Delta_Q^{-\sigma}e_{\sigma\epsilon}.
\end{aligned}
\label{eq:normalized-radiative-fields}
\end{equation}
The coefficients $g_{\sigma\epsilon},e_{\sigma\epsilon}$, listed in Eq.~\eqref{eq:reconstruction-coefficients}, include the complete mode normalization; no subsequent rescaling is applied.
For example,
\begin{equation}
\begin{aligned}
 g_{++}&=MC_\varrho w\chi
 -\frac{q}{3\varrho}\left(\frac{\Delta_Q}{M}w_\flat\chi
                         +\ii A_\theta w\zeta\right),\\
 e_{+-}&=MC_\varrho w_\flat\chi
 +\frac{q}{3\varrho}(Mw\chi+\ii A_\theta w_\flat\zeta).
\end{aligned}
\end{equation}
All eight reconstruction coefficients have no inverse powers of $Q$.

At zero charge their surviving entries are $M^2W_\tau\chi_\epsilon$ and the other entries vanish.
Since $\mathcal B_s(0)=M^{2s-2}H_0/\sqrt{\sin\theta}$, this gives directly
\begin{equation}
\begin{aligned}
 (\mathcal G_\sigma,\mathcal E_\sigma)\big|_{Q=0}
 =\begin{cases}(R_sS_s,0),&|s|=2,\\(0,R_sS_s),&|s|=1.
 \end{cases}
\end{aligned}
\label{eq:unit-kerr-field}
\end{equation}
The NP meaning of this correspondence is
\begin{equation}
\begin{array}{c|c|c}
s&\text{surviving field}&\text{Kerr value }R_sS_s\\\hline
+2&\mathcal G_+&\delta\Psi_0\\
+1&\mathcal E_+&\delta\Phi_0\\
-2&\mathcal G_-&(\varrho^*)^4\delta\Psi_4\\
-1&\mathcal E_-&(\varrho^*)^2\delta\Phi_2
\end{array}.
\label{eq:physical-teukolsky-fields}
\end{equation}
Restoring $e^{-\ii\omega t+\ii m\phi}$ gives the standard separated Kerr fields.

At finite charge, $\mathcal E_\pm$ are mixed curvature--Maxwell invariants, and the reconstructed fields need not factor into a single product $R_sS_s$.
Their gravitational or electromagnetic branch is even in $Q$, while the opposite field is odd, for separated solution families regular in $q$.
Some gravitational-branch mixed invariants vanish as $Q^3$; the physical Maxwell amplitude instead requires the extraction in Sec.~\ref{sec:flux}.

The inverse map gives radiative fields with fixed normalization; it does not construct a metric and vector potential for every separated solution or establish the compatibility and reality conditions relating the positive and negative spin systems.

\section{Boundary normalization and connection}\label{sec:scope}

To compare scattering amplitudes across zero charge, the ingoing and outgoing solutions must be normalized at the endpoints, not only at a finite radius.
The regular first-order system already gives analytic local solutions along a bounded holomorphic $\mu(q)$ curve, as detailed in Appendix~\ref{app:radial-local}.
We construct an infinity basis with regular normalization, connect it to a horizon basis, and then translate both bases into the radial functions $R_s$.
The obstruction is that fixed-radius charge expansions are nonuniform at large radius.

\subsection{A nonuniform far region}

We expand the radial coupling coefficients~\eqref{eq:FG} exactly as
\begin{align}
\begin{aligned}
 F_Q&=F_1r+F_0,\\
 G_Q&=G_1r+G_0,\\
 F_1&=\frac{2\ii\sigma\omega q}{3(M^2-q)},\qquad
 F_0=\frac{1+L}{2}-\frac{qF_1}{M}, \\
 G_1&=-\frac{6\ii\sigma\omega M^2}{M^2-q},\qquad
 G_0=\frac{2\mu}{1+L}-\frac{qG_1}{M}.
\end{aligned}
\label{eq:FG-r-dep}
\end{align}
The radial coefficient $F_Q$ exposes a limit that cannot be interchanged without analysis.
For nonzero $\omega$,
$F_Q\to1$ for $Q\to0$ at fixed $r$, but its $Q^2\omega r$ term dominates at sufficiently large $r$ for every nonzero charge.
The crossover scale is of order $M^2/(|\omega|Q^2)$ when $|Q|\ll M$ and bounded $\mu$.
The transformation $W_\tau=H_Q^{-1}(\Delta_Q/M^2)^s R_s$ of Eq.~\eqref{eq:finite-teuk-variable} contains $e^{\ii\xi_Qr}$,
whose expansion in $Q^2$ is nonuniform on the same scale.

These facts do not prove a discontinuity of any physical scattering amplitude; they show that fixed-radius convergence cannot establish its continuity.
The first-order system is regular at a zero of the coefficient used to eliminate a component; such a denominator is not a physical pole of the connection problem.

\subsection{Normalized horizon and infinity bases}\label{sec:boundary-normalization}\label{sec:infinity}

Fix real $\omega\ne0$ and a bounded holomorphic separation curve $\mu(q)$ near zero charge, such as the curves in Sec.~\ref{sec:angular-branch}.
We first normalize two independent solutions at each endpoint in the fixed component order $Y=(w,w_\flat)^T$.
The radial system~\eqref{eq:regular-radial-pair-unified} is
\begin{equation}
 Y'=\begin{pmatrix}0&\mathcal P_\tau/M\\M\mathcal R_\tau/\Delta_Q&-B_Q/\Delta_Q\end{pmatrix}Y,
\label{eq:radial-Y-system}
\end{equation}
Let us consider a fundamental matrix, an invertible matrix whose columns are independent solutions of the first-order system; here it contains two solution columns.
We write the infinity and horizon fundamental matrices as
\begin{align}
 \mathsf I(r,q)&=\bigl(Y^{\mathrm I}_1(r,q),Y^{\mathrm I}_2(r,q)\bigr),
 \label{eq:infinity-matrix-definition} \\
 \mathsf H(r,q)&=\bigl(Y^{\mathrm H}_1(r,q),Y^{\mathrm H}_2(r,q)\bigr).
 \label{eq:horizon-matrix-definition}
\end{align}
Each $Y^{\mathrm I}_j$ or $Y^{\mathrm H}_j$ is a two-component solution in $(w,w_\flat)$ space.
The superscripts identify infinity and the horizon, respectively, while $j=1,2$ labels independent solutions, not their components.
Equations~\eqref{eq:infinity-basis} and \eqref{eq:horizon-basis} fix their respective normalizations.
The following prescriptions fix their columns completely, including constant phases and subleading additions.

At infinity, $\mathcal P_\tau$ and $\mathcal R_\tau$, which equal $F_Q$ and $G_Q$ in the branch-dependent order of Eq.~\eqref{eq:radial-coupling-coefficients-renamed}, are linear functions of $r$.
Rescaling the second component by $r/M$ therefore makes both off-diagonal coefficients in the transformed first-order matrix approach constants.
Removing the common phase without expanding it gives
\begin{equation}
 Z=e^{-\ii\xi_Qr}\operatorname{diag}(1,r/M)Y.
 \label{eq:infinity-state-map}
\end{equation}
The leading coefficient of the transformed equation $Z'=\mathsf B(r)Z$ in Eq.~\eqref{eq:infinity-matrix} at infinity has the finite eigenvector matrices
\begin{equation}
\begin{aligned}
 E_+&=\begin{pmatrix}1&\sigma q/(3M^2)\\3\sigma&1\end{pmatrix},\\
 E_-&=\begin{pmatrix}1&-3\sigma\\-\sigma q/(3M^2)&1\end{pmatrix},
\end{aligned}
\label{eq:infinity-diagonalization}
\end{equation}
Here, $\det E_\tau=1-q/M^2$ is nonzero near zero charge.
The corresponding solution columns have exponential rates $0$ and $2\ii\omega$ and powers $(r/M)^{3\sigma}$ and $(r/M)^{4\ii M\omega}$, respectively.
These rates and powers are independent of $q$; the common phase $e^{\ii\xi_Qr}$ is restored exactly in $Y$.
Appendix~\ref{app:infinity-proof} gives the coefficient matrix and derives these statements.

Leading asymptotics alone may leave freedom to add a subdominant solution to a column.
We fix $\mathsf I$ by the exact factorization~\eqref{eq:infinity-basis}, with the specified near-identity transformation $\mathsf S$ and the residual condition $\mathsf U_\infty(\infty)=\Id$ in Eq.~\eqref{eq:infinity-volterra}.
That prescription defines exactly the matrix in Eq.~\eqref{eq:infinity-matrix-definition}, with no free constant right factor.
Its columns depend holomorphically on $q$ at a fixed exterior point.

At the horizon, using $d=r_+-r_-$ from Eq.~\eqref{eq:horizon-coordinates}, define
\begin{equation}
 \nu_\pm=\frac{3\sigma-1}{2}\pm\frac{2\ii K_0(r_\pm)}d.
 \label{eq:horizon-exponents}
\end{equation}
The local Frobenius exponents are $0$ and $\nu_+$.
We assume $\nu_+\notin\mathbb Z$, avoiding integer exponent differences at which the Frobenius recurrence can require logarithmic terms or a separate normalization analysis.
For real frequency on a subextremal background, this means $k_\mathrm{H}=\omega-m\Omega_\mathrm{H}\ne0$.
The excluded mode is stationary along the corotating horizon generator $\partial_t+\Omega_\mathrm{H}\partial_\phi$; this local condition is distinct from the global QNM boundary condition.
The two pure Frobenius solutions with leading vectors fixed in Eq.~\eqref{eq:horizon-basis} define $\mathsf H$ and depend holomorphically on $q$ under this assumption.
Appendix~\ref{app:horizon-proof} supplies their construction.

\subsection{Normalized radial waves and their Kerr limit}\label{sec:finite-boundaries}\label{sec:kerr-amplitudes}

We now convert the two-component bases into four radial waves using the variable $R_s$ fixed in Eq.~\eqref{eq:finite-teuk-variable}.
The horizon waves will have unit ingoing or outgoing amplitude.
At infinity we choose unit Kerr amplitudes and a regular continuation through $Q=0$; unit leading amplitudes at finite charge are a different convention, related explicitly below.
In addition to the preceding real-frequency and nonresonance assumptions, take the charge sufficiently small that $F_Q(r_+)\ne0$.

For any solution $Y$, the radial map~\eqref{eq:finite-teuk-variable} is
\begin{equation}
\begin{aligned}
 R_s[Y]&=H_Q(\Delta_Q/M^2)^{-s}e_\tau^T Y,\\
 e_+^T&=(1,0),\qquad e_-^T=(0,1).
\end{aligned}
\label{eq:finite-canonical-map}
\end{equation}
The radial map selects $w$ or $w_\flat$ and multiplies it by the displayed prefactor without another normalization.
With the tortoise coordinate fixed by Eq.~\eqref{eq:KN-tortoise}, the desired horizon waves obey
\begin{equation}
 \begin{aligned}
 R_s^{\mathrm{in}}&\sim(\Delta_Q/M^2)^{-s}e^{-\ii k_\mathrm{H}r_*},\\
 R_s^{\mathrm{out}}&\sim e^{\ii k_\mathrm{H}r_*},
 \end{aligned}
 \label{eq:finite-horizon-radial}
\end{equation}
Here, $\sim$ means that the ratio tends to one.
In the dimensionful convention $\Delta_Q^{-s}$, the unit ingoing solution is instead $M^{-2s}R_s^{\mathrm{in}}$.

Comparing these waves with the normalized columns of $\mathsf H$, and fixing the infinity columns to their unit Kerr values, gives the following diagonal conversion factors.
Define $F_+=F_Q(r_+)$, $B_+=B_Q(r_+)$, and the finite part of the horizon tortoise coordinate,
\begin{equation}
 r_*^{\mathrm H}=r_+-\frac{r_-^2+a^2}{d}\log(d/M).
 \label{eq:horizon-tortoise-constant}
\end{equation}
Thus, $r_*=(r_+^2+a^2)\log x/d+r_*^{\mathrm H}+\order{x}$ near the horizon.
The conversion factors are
\begin{equation}
\begin{aligned}
 D_\mathrm H={}&e^{-\ii(\omega+\xi_Q)r_+}(d/M)^{-\ii\alpha_-}\\
 &\times\operatorname{diag}\bigl(e^{\ii k_\mathrm H r_*^{\mathrm H}},
 (d/M)^{-s}e^{-\ii k_\mathrm H r_*^{\mathrm H}}\bigr)\\
 &\times\begin{cases}
 \operatorname{diag}(1,F_+/(\nu_++1)),&\tau=+1,\\
 \operatorname{diag}(MF_+/B_+,1),&\tau=-1,
 \end{cases}\\
 D_\mathrm I={}&\begin{cases}
 \operatorname{diag}(1,(2\ii M\omega)^{-1}),&\tau=+1,\\
 \operatorname{diag}(-(2\ii M\omega)^{-1},1),&\tau=-1.
 \end{cases}
\end{aligned}
\label{eq:boundary-normalization-factors}
\end{equation}
The retained Frobenius components and the subleading Kerr components used in this comparison are given in Eqs.~\eqref{eq:horizon-retained-components} and \eqref{eq:kerr-subleading-components}.
Nonresonance excludes $\nu_+=-1$, and $\operatorname{Re}B_+=(1-3\sigma)d/2\ne0$ for real parameters, so these denominators introduce no additional exclusions.
The four normalized solutions are
\begin{equation}
\begin{aligned}
 (R_s^{\mathrm{in}},R_s^{\mathrm{out}})
   &=R_s[\mathsf H]D_\mathrm{H}^{-1},\\
 (R_s^{\mathrm{down}},R_s^{\mathrm{up}})
   &=R_s[\mathsf I]D_\mathrm{I}^{-1},
\end{aligned}
\label{eq:normalized-four-solutions}
\end{equation}
where, for a matrix with solution columns, the map is the row vector
\begin{equation}
 R_s[(Y_1,Y_2)]\coloneqq\bigl(R_s[Y_1],R_s[Y_2]\bigr).
 \label{eq:radial-map-columns}
\end{equation}
The labels in/out refer to the horizon, and down/up to incoming/outgoing waves at infinity.
This fixes the radial-wave amplitudes; their physical energy fluxes require the reconstruction and radiation formulas below.

At infinity, let $\rho=r/M$ and $e_\pm^\infty=e^{\pm\ii\omega r}\rho^{\pm2\ii M\omega}$.
The large-radius expansion gives
\begin{equation}
\begin{aligned}
 R_s^{\mathrm{down}}&=\rho^{-1+\bar\tau}e_-^\infty
       \bigl[A_\tau(q)+A_\tau^{(1)}(q)/\rho+\order{\rho^{-2}}\bigr],\\
 R_s^{\mathrm{up}}&=\rho^{-2s-\bar\tau}e_+^\infty
       \bigl[B_\tau(q)+B_\tau^{(1)}(q)/\rho+\order{\rho^{-2}}\bigr],\\
 A_+(q)&=B_-(q)=1,\\
 A_-(q)&=B_+(q)=\frac{2\ii\sigma\omega q}{3M}.
\end{aligned}
\label{eq:finite-infinity-radial}
\end{equation}
The coefficients $A_-(q)$ and $B_+(q)$ vanish as $q\to0$ because one leading component has zero projection onto the Kerr wave.
The nonzero Kerr solution comes from the next power in $1/r$, and $A_-^{(1)}(0)=B_+^{(1)}(0)=1$ as shown in Appendix~\ref{app:kerr-infinity-normalization}.
With our normalization, direct evaluation at $q=0$ gives
\begin{equation}
\begin{aligned}
 R_s^{\mathrm{down}}(0,r)&\sim\rho^{-1}e^{-\ii\omega r_*},\\
 R_s^{\mathrm{up}}(0,r)&\sim\rho^{-2s-1}e^{\ii\omega r_*}.
\end{aligned}
\label{eq:unit-kerr-boundary}
\end{equation}
Both ratios tend to one; the horizon limit follows directly from Eq.~\eqref{eq:finite-horizon-radial}.
For the dimensionful Kerr conventions $r^{-1}e^{-\ii\omega r_*}$ and $r^{-2s-1}e^{\ii\omega r_*}$, the unit solutions are $M^{-1}R_s^{\mathrm{down}}$ and $M^{-2s-1}R_s^{\mathrm{up}}$, respectively.

For $q\ne0$, one may instead impose unit leading coefficients in the finite-charge powers by
\begin{equation}
 R_{s,\mathrm{unit}}^{\mathrm{down}}=\frac{R_s^{\mathrm{down}}}{A_\tau(q)},\qquad
 R_{s,\mathrm{unit}}^{\mathrm{up}}=\frac{R_s^{\mathrm{up}}}{B_\tau(q)}.
\label{eq:finite-unit-infinity}
\end{equation}
This conversion divides by $q$ in one channel of each branch and is singular at zero charge.
For example, $R_s^{\mathrm{up}}/B_+(q)$ diverges at fixed radius although its leading infinity coefficient becomes one.
Thus, unit finite-charge leading coefficients and a nonzero regular Kerr limit cannot both be imposed on these same branches.
Equation~\eqref{eq:normalized-four-solutions} is our primary convention; the finite-charge conversion above is optional and restricted to $q\ne0$.

\subsection{Connection coefficients and outgoing amplitudes}\label{sec:connection-amplitudes}

The normalized endpoint bases determine the connection matrix
\begin{equation}
 \mathsf H(r,q)=\mathsf I(r,q)C(q).
 \label{eq:connection-definition}
\end{equation}
Thus, $C(q)=\mathsf I(r_b,q)^{-1}\mathsf H(r_b,q)$ is independent of the exterior matching radius $r_b$ and holomorphic near zero charge under the assumptions above.
This is the boundary-normalized counterpart of local regularity; it does not imply unweighted uniform convergence of the original variables throughout the exterior, where $e^{\ii\xi_Qr}$ remains nonuniform.
Abel's identity and the endpoint normalizations give
\begin{equation}
 \det C=\frac{e^{-2\ii(\omega+\xi_Q)r_+}}{1-q/M^2}
 \left(\frac{M}{d}\right)^{\nu_-}.
 \label{eq:connection-determinant}
\end{equation}
The determinant has a finite nonzero Kerr limit and checks the relative normalization.
Individual matrix entries can still vanish, so ratios of connection coefficients require separate attention at such zeros.
Appendix~\ref{app:connection-proof} gives the proof and a numerical check with prescribed $\mu$.

For the normalized radial waves, the corresponding connection is
\begin{equation}
\begin{aligned}
 (R_s^{\mathrm{in}},R_s^{\mathrm{out}})
  &=(R_s^{\mathrm{down}},R_s^{\mathrm{up}})C_T(q),\\
 C_T(q)&=D_\mathrm{I}C(q)D_\mathrm{H}(q)^{-1}.
\end{aligned}
\label{eq:kerr-amplitude-dictionary}
\end{equation}
For example, $R_s^{\mathrm{in}}=(C_T)_{11}R_s^{\mathrm{down}}+(C_T)_{21}R_s^{\mathrm{up}}$, with no unspecified proportionality factor.
If unit finite-charge leading amplitudes are used instead, Eq.~\eqref{eq:finite-unit-infinity} gives the connection matrix $\operatorname{diag}(A_\tau,B_\tau)C_T$.

To obtain the radiation of a solution specified by arbitrary boundary data, expand it in the normalized infinity waves:
\begin{equation}
 R_s=b_\mathrm{down}R_s^{\mathrm{down}}+b_\mathrm{up}R_s^{\mathrm{up}}.
\end{equation}
The constants $b_\mathrm{down},b_\mathrm{up}$ are expansion coefficients in this solution basis.
For instance, for $R_s=R_s^{\mathrm{in}}$, $b_\mathrm{up}=(C_T)_{21}$.
For general $b_\mathrm{down},b_\mathrm{up}$, the corresponding state and master outgoing coefficient are
\begin{equation}
\begin{aligned}
 Y&=\frac{b_\mathrm{down}}{(D_\mathrm I)_{11}}Y_1^{\mathrm I}
   +\frac{b_\mathrm{up}}{(D_\mathrm I)_{22}}Y_2^{\mathrm I},\\
 A_\mathrm{out}&\coloneqq\frac{b_\mathrm{up}}{(D_\mathrm I)_{22}}.
\end{aligned}
\label{eq:boundary-to-master-amplitude}
\end{equation}
Thus, $A_\mathrm{out}$ multiplies the fully normalized outgoing column $Y_2^{\mathrm I}$ and carries the entire radial amplitude entering Sec.~\ref{sec:flux}.

The boundary construction concerns real nonzero frequency and a nonresonant horizon.
For complex $\omega$, exponential growth can invalidate the real-ray Volterra estimate; QNM poles and residues require a separate continuation, together with angular spectral control.

\section{Outgoing radiation and energy flux}\label{sec:flux}

The regular radiative invariants determine the energy carried to future null infinity once the asymptotic frame is fixed.
In this section, we consider nonzero real $\omega$ and the outgoing part of a physical Einstein--Maxwell perturbation with the usual peeling expansion in a regularly perturbed Kinnersley frame~\cite{Newman:1961qr,Penrose:1965am}.
Throughout this section, $\sigma=-1$, so $\epsilon=+1$ corresponds to the gravitational branch $s=-2$, and $\epsilon=-1$ to the electromagnetic branch $s=-1$.

The physical-origin assumption means that the radiative fields arise from a linearized Einstein--Maxwell solution, so that the asymptotic Bianchi identity can be imposed on them.
The amplitude and flux statements in this section are conditional on this assumption; inversion of the separated radiative subsystem alone does not establish it for an arbitrary master solution.
Appendix~\ref{app:flux-conventions} derives the asymptotic Bianchi identity and flux factors used below.
The gravitational amplitude follows directly from $\mathcal G_-$, but $\mathcal E_-$ also contains a gravitational contribution through $\delta\Psi_3$.
We extract the leading asymptotic coefficients, use the Bianchi identity to subtract that contribution, and then evaluate the two physical fluxes.

Let $u=t-r_*=t-r-2M\log(r/M)+\order{r^{-1}}$, with $r_*$ fixed by Eq.~\eqref{eq:KN-tortoise}.
Suppressing $e^{-\ii\omega u+\ii m\phi}$, define
\begin{align}
 \delta\Psi_4&=g(\theta)/r+\order{r^{-2}},\\
 \delta\Psi_3&=j(\theta)/r^2+\order{r^{-3}},\\
 \delta\Phi_2&=f(\theta)/r+\order{r^{-2}},\\
 \mathcal E_-&=e(\theta)r+\order{1}.
\end{align}
These amplitudes follow directly from the reconstructed fields:
\begin{equation}
\begin{aligned}
 g(\theta)&=\lim_{r\to\infty}\frac{\mathcal B_s\Delta_Q^2g_{-\epsilon}}{r^3},\\
 e(\theta)&=\lim_{r\to\infty}\frac{\mathcal B_s\Delta_Qe_{-\epsilon}}r.
\end{aligned}
\label{eq:reconstruction-to-radiation}
\end{equation}
The outgoing Fourier factor is removed before taking these limits.
Thus, $g_{\sigma\epsilon},e_{\sigma\epsilon}$ describe the full separated reconstruction, whereas $g(\theta),e(\theta)$ describe its leading radiation at infinity.
The background $\Psi_2^{(0)}=-M/r^3+\order{r^{-4}}$ and Eq.~\eqref{eq:regular-radiation} give
\begin{equation}
 e(\theta)=f(\theta)+\frac{Q}{3M}j(\theta).
 \label{eq:asymptotic-mixed}
\end{equation}
Both contributions on the right have the same radial order.
The mixed invariant cannot therefore be identified directly with the Maxwell radiation amplitude.

\subsection{Removing the gravitational contribution}

In the adopted tetrad, the leading nonradial Bianchi identity~\eqref{eq:NP-bianchi-full} yields
\begin{equation}
\begin{aligned}
 -\ii\omega j(\theta)&=\frac1{\sqrt2}\mathscr L_2^\dagger g(\theta),\\
 \mathscr L_2^\dagger&=\partial_\theta-m\csc\theta
 +a\omega\sin\theta+2\cot\theta.
\end{aligned}
\label{eq:radiative-bianchi}
\end{equation}
Appendix~\ref{app:flux-conventions} derives the factors and states the asymptotic assumptions.
In particular, the $a\omega\sin\theta$ term comes from the time component of the Kinnersley angular leg; replacing $\mathscr L_2^\dagger$ by a spherical angular derivative would mix frame conventions.
Combining Eqs.~\eqref{eq:asymptotic-mixed} and \eqref{eq:radiative-bianchi} gives the physical Maxwell amplitude
\begin{equation}
 f(\theta)=e(\theta)-\frac{\ii Q}{3\sqrt2M\omega}\mathscr L_2^\dagger g(\theta).
 \label{eq:physical-Maxwell-extraction}
\end{equation}
This extraction uses the leading asymptotic field equations and does not require a full metric reconstruction in the interior.
At zero frequency the inversion of $\partial_u$ is unavailable, so Eq.~\eqref{eq:physical-Maxwell-extraction} is not asserted there.

\subsection{Physical amplitudes of a separated outgoing mode}\label{sec:physical-outgoing}

Let $A_{\rm out}$ multiply the outgoing column $Y_2^{\mathrm I}$ of the normalized master basis.
For a solution specified by its radial-wave coefficient $b_{\rm up}$, Eq.~\eqref{eq:boundary-to-master-amplitude} fixes $A_{\rm out}$ with no further phase or normalization choice.
For $\sigma=-1$, the final physical amplitudes are
\begin{equation}
\begin{aligned}
 g(\theta)&=\frac{A_{\rm out}}{\sqrt{\sin\theta}}
 \begin{cases}\chi_+/M^3,&s=-2,\\
 Q\zeta_-/(3\sqrt2M^3),&s=-1,\end{cases}\\
 f(\theta)&=-\frac{\ii A_{\rm out}}{\sqrt{\sin\theta}}
 \begin{cases}Q\zeta_+/(3\sqrt2M^4\omega),&s=-2,\\
 \chi_-/(2M^2\omega),&s=-1.\end{cases}
\end{aligned}
\label{eq:physical-outgoing-amplitudes}
\end{equation}
These expressions are exact at finite rotation and charge for nonzero real $\omega$ on the spectral branch $L(0)=1$ with $1+L\ne0$, subject to the physical-origin and peeling assumptions above.
All dependence on the radial excitation, including its phase and normalization, is carried by $A_{\rm out}$; rotation also enters the angular functions.

To obtain Eq.~\eqref{eq:physical-outgoing-amplitudes}, insert the outgoing radial solution into the normalized field inverse of Sec.~\ref{sec:physical}.
Its leading eigenvector determines $g(\theta)$, but the leading contributions to the mixed electromagnetic invariant cancel, requiring one further radial order.
Subtracting the gravitational contribution according to Eq.~\eqref{eq:physical-Maxwell-extraction} and using the first-order angular equations then cancels the explicit rotation-dependent terms.
Appendix~\ref{app:outgoing-profiles} gives the intermediate profiles and both cancellations.
In particular, the gravitational branch has a physical Maxwell amplitude of order $Q/M$, even though its mixed electromagnetic profile begins at order $(Q/M)^3$.

\subsection{Energy carried to infinity}

Let $\mathcal F_\mathrm{G}$ and $\mathcal F_\mathrm{E}$ denote the energy radiated per unit retarded time and solid angle in the gravitational and electromagnetic channels, respectively.
They are the large-radius limits of $r^2$ times the local radial energy flux, rather than flux per unit area at finite radius.
In the adopted Kinnersley normalization, these angular power densities are
\begin{equation}
 \mathcal F_\mathrm{G}=\frac{|g(\theta)|^2}{4\pi\omega^2},\qquad
 \mathcal F_\mathrm{E}=\frac{|f(\theta)|^2}{2\pi}.
 \label{eq:flux-normalization}
\end{equation}
These formulas use the physical spin-complex radiation data: the two real polarizations form the real and imaginary components.
They also apply to a consistently normalized monochromatic polarization mode; a different real-part/time-average convention must be converted before comparing amplitudes.
Appendix~\ref{app:flux-conventions} checks the factors directly with asymptotic plane waves.
Using Eq.~\eqref{eq:physical-Maxwell-extraction}, the total flux is consequently
\begin{equation}
 \mathcal F_\infty=
 \frac{|g(\theta)|^2}{4\pi\omega^2}
 +\frac1{2\pi}
 \left|e(\theta)-\frac{\ii Q}{3\sqrt2M\omega}\mathscr L_2^\dagger g(\theta)\right|^2.
 \label{eq:physical-flux}
\end{equation}
Integrating over $\dd\Omega=\sin\theta\,\dd\theta\,\dd\phi$ defines the outgoing powers,
\begin{equation}
 P_\mathrm{G}=\int\mathcal F_\mathrm{G}\,\dd\Omega,\qquad
 P_\mathrm{E}=\int\mathcal F_\mathrm{E}\,\dd\Omega.
 \label{eq:integrated-powers}
\end{equation}
For a superposition, one first sums the physical amplitudes $g,f$ and then forms their quadratic fluxes; orthogonality of distinct finite-charge separation branches is not assumed.
Equations~\eqref{eq:physical-outgoing-amplitudes} and \eqref{eq:physical-flux} convert the normalized master outgoing coefficient into physical radiation.
The two nonnegative terms are separate physical channels; interference inside the second term reflects the mixed choice of master invariant.

For a single real-frequency mode $e^{-\ii\omega t+\ii m\phi}$, the outgoing axial angular-momentum flux is also fixed:
\begin{equation}
 \frac{\dd J_z}{\dd t}=\frac{m}{\omega}(P_\mathrm{G}+P_\mathrm{E}).
 \label{eq:angular-momentum-flux}
\end{equation}

The ratio follows from the time-translation and axial-rotation generators acting on the same radiative mode, as in the Kerr relation of Ref.~\cite{Teukolsky:1974yv}.
Here, $P_\mathrm{G},P_\mathrm{E}$ are its integrated outgoing powers; the same polarization and averaging convention must be used for both fluxes.

\subsection{Exact radiation composition and its small-charge limit}

For one separated outgoing mode, the physical powers now require only the norms of the two regular angular components.
Define
\begin{equation}
 N_{\chi,\epsilon}=\int_0^\pi|\chi_\epsilon|^2\,\dd\theta,
 \qquad
 N_{\zeta,\epsilon}=\int_0^\pi|\zeta_\epsilon|^2\,\dd\theta.
\end{equation}
The azimuthal integral contributes $2\pi$.
Using Eq.~\eqref{eq:boundary-to-master-amplitude}, namely $A_{\rm out}=b_{\rm up}$ for $\epsilon=+$ and $A_{\rm out}=2\ii M\omega b_{\rm up}$ for $\epsilon=-$, gives
\begin{equation}
\begin{aligned}
 (P_\mathrm{G},P_\mathrm{E})_{s=-2}&=\frac{|b_{\rm up}|^2}{2M^6\omega^2}
 \left(N_{\chi,+},\frac{Q^2}{9M^2}N_{\zeta,+}\right),\\
 (P_\mathrm{G},P_\mathrm{E})_{s=-1}&=\frac{|b_{\rm up}|^2}{M^2}
 \left(\frac{Q^2}{9M^2}N_{\zeta,-},N_{\chi,-}\right).
\end{aligned}
\label{eq:boundary-to-power}
\end{equation}
Each line uses the angular pair and outgoing radial coefficient of that branch.
Their normalizations must be consistent: rescaling $(\chi,\zeta)$ by a nonzero constant and $b_{\rm up}$ by its reciprocal leaves the physical powers unchanged.
For the horizon-normalized solution $R_s^{\rm in}$, insert $b_{\rm up}=(C_T)_{21}$.
This evaluates its outgoing power; no normalization by an incident physical flux is implied.
The amplitude-independent composition ratios are therefore
\begin{equation}
\begin{aligned}
 \left.\frac{P_\mathrm{E}}{P_\mathrm{G}}\right|_{s=-2}
 &=\frac{Q^2}{9M^2}\frac{N_{\zeta,+}}{N_{\chi,+}},\\
 \left.\frac{P_\mathrm{G}}{P_\mathrm{E}}\right|_{s=-1}
 &=\frac{Q^2}{9M^2}\frac{N_{\zeta,-}}{N_{\chi,-}}.
\end{aligned}
\label{eq:exact-radiation-ratios}
\end{equation}
These are exact finite-charge results for a single mode with nonzero denominator power.
The common radial amplitude cancels, so the composition of a single outgoing mode depends only on its angular branch, independently of the radial excitation or horizon boundary data.
The branch label identifies its Kerr origin, not which channel must dominate at arbitrary finite charge.
These are outgoing radiation-composition ratios, not scattering conversion probabilities.

The exact amplitudes also give a rotating small-charge prediction using only Kerr angular functions.
Consider the local separated branch constructed in Sec.~\ref{sec:angular-branch} at fixed real $\omega\ne0$.
Its pair $(\chi_\epsilon,\zeta_\epsilon)$ is holomorphic in $L^2$, which controls the angular integrals below without requiring $\chi_\epsilon'$ itself to be square integrable at a critical pole.
Let $S_{-2}$ or $S_{-1}$ be its limiting Kerr angular function.
Define $\|S\|_\Omega^2=2\pi\int_0^\pi|S|^2\sin\theta\,\dd\theta$ and
\begin{equation}
 \mathscr L_{-1}=\partial_\theta+m\csc\theta
 -a\omega\sin\theta-\cot\theta.
\end{equation}
For the gravitational branch, Eq.~\eqref{eq:physical-outgoing-amplitudes} gives $g(\theta)=(A_{\rm out}/M^3)[S_{-2}+\order{q/M^2}]$ and a Maxwell amplitude of order $Q/M$.
At $q=0$, Eq.~\eqref{eq:angular-division-free} gives $\zeta_\epsilon=(\partial_\theta+\epsilon t_0)\chi_\epsilon$ because $\beta_0=1$.
Substituting $\chi_\epsilon=\sqrt{\sin\theta}S_s$ at $\sigma=-1$ gives
\begin{equation}
 \frac{\zeta_+}{\sqrt{\sin\theta}}=\mathscr L_2^\dagger S_{-2},\qquad
 \frac{\zeta_-}{\sqrt{\sin\theta}}=\mathscr L_{-1}S_{-1}.
\end{equation}
Along the local charge-dependent branch these relations receive $\order{q/M^2}$ corrections.
The electromagnetic amplitude divided by $A_{\rm out}$ is therefore $-\ii S_{-1}/(2M^2\omega)+\order{q/M^2}$.
Consequently,
\begin{equation}
\begin{aligned}
 \left.\frac{P_\mathrm{E}}{P_\mathrm{G}}\right|_{s=-2}
 &=\frac{Q^2}{9M^2}
 \frac{\|\mathscr L_2^\dagger S_{-2}\|_\Omega^2}
 {\|S_{-2}\|_\Omega^2}+\order{(Q/M)^4},\\
 \left.\frac{P_\mathrm{G}}{P_\mathrm{E}}\right|_{s=-1}
 &=\frac{Q^2}{9M^2}
 \frac{\|\mathscr L_{-1}S_{-1}\|_\Omega^2}
 {\|S_{-1}\|_\Omega^2}+\order{(Q/M)^4}.
\end{aligned}
\label{eq:small-charge-radiation-ratios}
\end{equation}
Here, the remainder notation holds at fixed $M,a,\omega$ and fixed angular branch.
No finite-charge angular eigenvalue derivative enters the leading coefficient, since it would first correct the minority-channel power at order $Q^4$.

The Kerr angular equation also removes the angular derivatives from these leading coefficients.
With
\begin{equation}
 \langle\cos\theta\rangle_s=
 \frac{\int |S_s|^2\cos\theta\,\dd\Omega}{\|S_s\|_\Omega^2},
\end{equation}
both derivative-norm ratios in Eq.~\eqref{eq:small-charge-radiation-ratios} equal
\begin{equation}
 \mu_0+6a\omega\langle\cos\theta\rangle_s,
 \qquad s=-2\ \text{or}\ -1,
 \label{eq:mixing-angular-moment}
\end{equation}
where $\mu_0$ is the corresponding Kerr separation parameter in Eq.~\eqref{eq:full-dictionary-identity}.
For any real coefficient function $v(\theta)$, the adjoint of $\partial_\theta+v(\theta)$ in the $\sin\theta$ measure is $-\partial_\theta-\cot\theta+v(\theta)$.
Applying the resulting second-order products to $S_s$ and using Eq.~\eqref{eq:teuk-angular} gives $\mu_0+6a\omega\cos\theta$.
Regularity makes the integration-by-parts boundary term vanish, including the poles with zero spin-weighted exponent, where the singular derivative combination cancels.

Thus, a Kerr eigenvalue and one angular moment suffice to evaluate the leading composition ratio without differentiating $S_s$.
Using Eq.~\eqref{eq:full-dictionary-identity} to eliminate $\mu_0$, Eq.~\eqref{eq:small-charge-radiation-ratios} gives
\begin{equation}
\begin{aligned}
 &\begin{cases}
 \left.P_\mathrm{E}/P_\mathrm{G}\right|_{s=-2},&s=-2,\\
 \left.P_\mathrm{G}/P_\mathrm{E}\right|_{s=-1},&s=-1
 \end{cases}\\
 &=\frac{Q^2}{9M^2}\Bigl[{}_sA_{\ell m}(a\omega)+a^2\omega^2-2ma\omega+2s(s+2)\\
 &\qquad\qquad+6a\omega\langle\cos\theta\rangle_s\Bigr]
 +\order{(Q/M)^4}.
\end{aligned}
\label{eq:small-charge-radiation-ratios-2}
\end{equation}
The spin weight enters both the eigenvalue ${}_sA_{\ell m}$ and the eigenfunction used in $\langle\cos\theta\rangle_s$, as well as the explicit term $2s(s+2)$.
Thus, the same functional expression generally gives different order-$Q^2$ coefficients for $s=-2$ and $s=-1$ at nonzero rotation.
At $a\omega=0$, ${}_sA_{\ell m}=\ell(\ell+1)-s(s+1)$, and both bracketed expressions reduce to $\ell(\ell+1)-2$.

\subsection{An exact nonrotating check}

At $a=0$, regular angular modes have
$\mu=\mu_0=\ell(\ell+1)-2$ at every charge, and $\beta_Q=(1+L)/2$.
For $\ell\ge2$, both branches exist.
The general amplitudes~\eqref{eq:physical-outgoing-amplitudes} already give the RN amplitudes by setting $a=0$.
The regular angular equation and integration by parts imply
\begin{equation}
 \int_0^\pi|\zeta_\epsilon|^2\,\dd\theta
 =\frac{4\mu}{(1+L)^2}
 \int_0^\pi|\chi_\epsilon|^2\,\dd\theta.
\end{equation}
Writing $P_\mathrm{G},P_\mathrm{E}$ for the integrated powers gives
\begin{equation}
 \left.\frac{P_\mathrm{E}}{P_\mathrm{G}}\right|_{s=-2}
 =\left.\frac{P_\mathrm{G}}{P_\mathrm{E}}\right|_{s=-1}
 =\frac{L-1}{L+1}.
 \label{eq:RN-power-ratio}
\end{equation}
Both ratios behave as $Q^2\mu/(9M^2)+\order{(Q/M)^4}$ and vanish in the Schwarzschild limit $Q\to0$ at $a=0$.

An independent comparison is supplied by the usual RN radiative channels~\cite{OuldElHadj:2023anu}.
We use the physical-channel normalization in which the reflection matrix of Ref.~\cite{OuldElHadj:2023anu} acts on flux-normalized electromagnetic and gravitational amplitudes and its off-diagonal modulus squared is the conversion probability.
For $\ell\ge2$, these amplitudes are related to the two decoupled channels by an orthogonal rotation with angle $\psi$.
Using the channel coefficients in Eqs.~(6) and (12)--(15) of Ref.~\cite{OuldElHadj:2023anu},
\begin{equation}
 \begin{aligned}
 \cos^2\psi&=\frac{d_\ell+3M}{2d_\ell},\qquad
 \sin^2\psi=\frac{d_\ell-3M}{2d_\ell}, \\
 d_\ell&=\sqrt{9M^2+4Q^2(\ell-1)(\ell+2)}.
 \end{aligned}
\end{equation}
Each decoupled channel therefore has smaller-to-larger radiative power ratio $\tan^2\psi=(d_\ell-3M)/(d_\ell+3M)$.
Since $\mu=\mu_0=(\ell-1)(\ell+2)$  and $d_\ell=3ML$, this is precisely Eq.~\eqref{eq:RN-power-ratio}.
The agreement checks the outgoing radiation composition against the independently normalized RN channels, including the relative sign in the Bianchi subtraction in Eq.~\eqref{eq:physical-Maxwell-extraction}.

\section{Conclusion}\label{sec:conclusion}

We have established a charge-regular formulation of separated Kerr--Newman perturbations, using the phantom-gauge separation as the starting point.
We have clarified the finite-charge master equations, both in the first-order and second-order forms, maintaining their smooth Kerr limits.
We have derived an explicit, normalized map from separated master solutions to outgoing gravitational and electromagnetic radiation.
For physical Einstein--Maxwell perturbations with the stated peeling behavior, the map determines both radiation amplitudes and their energy fluxes from the master boundary coefficients, within conventions that recover the standard Kerr fields at zero charge.

The angular and radial first-order systems~\eqref{eq:angular-division-free} and \eqref{eq:regular-radial-pair-unified-D} have the same ordered structure and remain the defining equations where a second-order elimination coefficient vanishes.
Their explicit second-order forms~\eqref{eq:explicit-teuk-angular} and \eqref{eq:explicit-teuk-radial} describe the finite-charge branches with one separation-constant dictionary and recover all four Teukolsky equations in Kerr by direct substitution.
These differential equations, the regular first-order systems, and their Kerr correspondence hold for complex frequencies on their stated coefficient domains.
This follows from the algebraic separated equations, not from extending a real-frequency mode-stability theorem.
Our proofs of angular-branch and boundary-connection analyticity, and the stated energy-flux formulas, retain the real-frequency restrictions specified above.
The inverse transformation~\eqref{eq:normalized-radiative-fields} reconstructs the finite-charge curvature--Maxwell fields with fixed normalization; each surviving Kerr field is precisely $R_sS_s$.

The boundary analysis is essential to this correspondence because fixed-radius convergence is insufficient at infinity.
An adapted first-order basis has charge-independent exponential rates and power indices, and defines a normalized connection matrix locally holomorphic in $Q^2$ at nonzero real frequency, subject to angular simplicity and horizon nonresonance.
The canonical radial variable gives the four boundary solutions and amplitude dictionary in Eqs.~\eqref{eq:normalized-four-solutions}--\eqref{eq:boundary-to-master-amplitude}.
Their Kerr limits have unit radial amplitudes.
At finite charge one leading infinity coefficient vanishes as $Q^2$ in each branch, explaining why unit finite-charge leading amplitudes and a regular Kerr limit require different conventions.
This construction does not make the original variables uniformly convergent over the entire exterior.

For outgoing physical perturbations with the usual peeling expansion, Eq.~\eqref{eq:physical-outgoing-amplitudes} gives the gravitational and electromagnetic amplitudes, and Eq.~\eqref{eq:physical-flux} gives their energy flux.
The Maxwell extraction requires both a gravitational subtraction and the first subleading radial coefficient.
The angular first-order system cancels the explicit rotation-dependent terms in the Maxwell extraction, leaving the simple physical amplitudes~\eqref{eq:physical-outgoing-amplitudes}.
They give the direct boundary-to-power formula~\eqref{eq:boundary-to-power} and exact finite-charge composition ratios~\eqref{eq:exact-radiation-ratios}.
The leading small-charge composition formula~\eqref{eq:small-charge-radiation-ratios-2} requires only Kerr angular eigenvalues and moments, without derivatives of the finite-charge angular eigenvalue.
The nonrotating limit reproduces the independently normalized RN channel mixture, providing a check of the relative gravitational and electromagnetic normalization.

The regular systems and explicit radiation amplitudes provide a starting point for further calculations of the coupled gravitational and electromagnetic response.
Horizon absorption and full scattering conversion remain to be evaluated.
Green-function residues and excitation factors additionally require control of complex-frequency continuation and the angular spectrum beyond the local branches considered here.
Full metric and vector-potential reconstruction is a separate problem.
These extensions can build on the regular systems and explicit normalization formulas collected in Table~\ref{tab:results}.

\smallskip
\emph{Note added.---}While this manuscript was being finalized, Ref.~\cite{Destounis:2026ljd} appeared, reporting QNM calculations based on the separated Kerr--Newman equations.
Its Appendix~B independently obtains charge-regular angular and radial systems and their Kerr limits, overlapping with and consistent with the results of Secs.~\ref{sec:angular} and \ref{sec:radial}.
The finite separation parameters are related by $\mu=3\tilde\lambda+(Q^2/M^2)\tilde\lambda^2$, where $\tilde\lambda$ is defined in Ref.~\cite{Destounis:2026ljd}.
Our normalized physical-field and radiation formulas in Secs.~\ref{sec:physical}--\ref{sec:flux} provide complementary results.

\acknowledgments
This work was supported in part by JSPS KAKENHI Grant Nos.~JP22K03639 and JP26K21813.
ChatGPT and Claude were used to assist with symbolic calculations, consistency checks, and manuscript preparation.
The author formulated the research questions, directed the analysis, critically reviewed and revised the AI-assisted calculations and manuscript, and takes responsibility for the analysis and conclusions.

\appendix
\renewcommand{\theequation}{\Alph{section}.\arabic{equation}}
\begin{widetext}
\section{NP conventions and asymptotic flux normalization}\label{app:flux-conventions}

We collect the conventions needed for the field inverse and the asymptotic Bianchi identity, using the NP framework of Refs.~\cite{Newman:1961qr,Chandrasekhar:1985kt} and the normalization of Ref.~\cite{Hintz:2026jiw}.
\subsection{Tetrad, curvature, and NP conventions}

The tetrad is Eq.~\eqref{eq:tetrad}, with $e_1=\ell$, $e_2=n$, $e_3=m$, $e_4=\bar m$, $\ell\cdot n=-1$, and $m\cdot\bar m=1$.
Our curvature sign is fixed by
\begin{equation}
 R(X,Y)Z=\bigl([\nabla_X,\nabla_Y]-\nabla_{[X,Y]}\bigr)Z,
 \qquad R(e_c,e_d)e_b=R^a{}_{bcd}e_a.
 \label{eq:curvature-convention}
\end{equation}
Writing $C_{abcd}$ for the tetrad components of the all-lowered Weyl tensor and $F_{ab}=F(e_a,e_b)$ for the electromagnetic field strength $F=dA$, the NP components are
\begin{equation}
\begin{aligned}
 \Psi_0&=C_{1313},& \Psi_1&=C_{1213},& \Psi_2&=C_{1342},&
 \Psi_3&=C_{1242},& \Psi_4&=C_{2424},\\
 \Phi_0&=F_{13},& \Phi_1&=\tfrac12(F_{12}+F_{43}),& \Phi_2&=F_{42}.
\end{aligned}
\label{eq:NP-component-conventions}
\end{equation}

Relative to the $(+---)$ convention of Ref.~\cite{Hintz:2026jiw}, the connection and $R^a{}_{bcd}$ are unchanged by the overall metric-sign reversal, whereas $C_{abcd}$ changes sign.
The plus signs in Eq.~\eqref{eq:NP-component-conventions} therefore preserve the numerical Weyl scalars; the Maxwell definitions are unchanged.
In particular, these definitions give the Coulomb fields~\eqref{eq:background-NP} and the phantom-gauge variables~\eqref{eq:source-NP-variables}.

To fix the spin-coefficient conventions without an implicit index-order choice, define
\begin{equation}
 \Gamma_{ijk}=g(\nabla_{e_k}e_i,e_j).
 \label{eq:tetrad-connection-convention}
\end{equation}
The spin coefficients used in this paper are
\begin{equation}
\begin{aligned}
 \kappa_{\rm NP}&=\Gamma_{311},&\sigma_{\rm NP}&=\Gamma_{313},&\lambda_{\rm NP}&=-\Gamma_{424},&\nu_{\rm NP}&=-\Gamma_{422},\\
 \beta_{\rm NP}&=\tfrac12(\Gamma_{343}-\Gamma_{123}),&\tau_{\rm NP}&=\Gamma_{312},&\mu_{\rm NP}&=-\Gamma_{423},&\gamma_{\rm NP}&=\tfrac12(\Gamma_{342}-\Gamma_{122}).
\end{aligned}
\label{eq:NP-spin-conventions}
\end{equation}
The first row vanishes on the aligned Kerr--Newman background; its perturbations enter the transformation to separated variables in Appendix~\ref{app:physical}.
The asymptotic Bianchi identity uses $\beta_{\rm NP},\tau_{\rm NP},\mu_{\rm NP},\gamma_{\rm NP}$ and $\delta\nu_{\rm NP}$.
Our $\Gamma_{ijk}$ is the negative of Hintz's Ricci rotation coefficient with the same index order, so these combinations retain the numerical NP coefficients.
For the background tetrad they evaluate to
\begin{equation}
\begin{aligned}
 \beta_{\rm NP}&=\frac{\cot\theta}{2\sqrt2\varrho},&
 \tau_{\rm NP}&=-\frac{\ii a\sin\theta}{\sqrt2\Sigma},&
 \mu_{\rm NP}&=-\frac{\Delta_Q}{2\Sigma\varrho^*},\\
 \gamma_{\rm NP}&=\mu_{\rm NP}+\frac{r-M}{2\Sigma},&
 \nu_{\rm NP}&=0.
\end{aligned}
\label{eq:NP-background-coefficients}
\end{equation}
At $Q=0$ these agree with the Kinnersley values in Ref.~\cite{Teukolsky:1973ha}.
The subscript ${\rm NP}$ distinguishes these coefficients from the separation parameter $\mu$, the branch sign $\tau$, and the angular couplings $\beta_Q,\gamma_Q$.
The following calculation uses the fixed spin and boost normalization of Eq.~\eqref{eq:tetrad}; invariance of the radiative combinations under the infinitesimal null rotations of Sec.~\ref{sec:review} does not remove this normalization choice.

\subsection{Asymptotic Bianchi identity}

The adopted Kinnersley tetrad and the usual NP Bianchi equation fix the sign and angular operator in Eq.~\eqref{eq:radiative-bianchi}~\cite{Hintz:2026jiw,Teukolsky:1973ha}.
Using the retarded time $u=t-r_*=t-r-2M\log(r/M)+\order{r^{-1}}$ of Sec.~\ref{sec:flux}, the leading action on outgoing fields is
\begin{align}
 n^a\partial_a&=\partial_u+\order{r^{-1}},\\
 m^a\partial_a&=\frac{\partial_\theta+\ii\csc\theta\partial_\phi+
 \ii a\sin\theta\partial_u}{\sqrt2 r}+\order{r^{-2}},\\
 \beta_{\rm NP}&=\frac{\cot\theta}{2\sqrt2 r}+\order{r^{-2}},\qquad
 \tau_{\rm NP}=\order{r^{-2}}.
\end{align}

The first line concerns the leading coefficient of an outgoing field; radial derivatives of its inverse-radius prefactor are subleading.
The relevant identity in the adopted convention is
\begin{equation}
 n^a\partial_a\delta\Psi_3-m^a\partial_a\delta\Psi_4
 =(4\beta_{\rm NP}-\tau_{\rm NP})\delta\Psi_4
 -2(2\mu_{\rm NP}+\gamma_{\rm NP})\delta\Psi_3
 +3\delta\nu_{\rm NP}\Psi_2^{(0)}+\text{linear Ricci terms}.
\label{eq:NP-bianchi-full}
\end{equation}
For a regular asymptotic frame and the stated peeling behavior, all terms except $(m^a\partial_a+4\beta_{\rm NP})\delta\Psi_4$ are subleading to $r^{-2}$ on the right.
In particular, linear electromagnetic Ricci terms involve the Coulomb background $\Phi_1^{(0)}=\order{r^{-2}}$ times radiation $\delta\Phi_2=\order{r^{-1}}$, while the radiative $\Phi_{22}$ stress is quadratic in perturbation amplitude.
Write $\delta\Psi_4=g(u,\theta,\phi)/r+\order{r^{-2}}$ and $\delta\Psi_3=j(u,\theta,\phi)/r^2+\order{r^{-3}}$; their Fourier amplitudes are the $g(\theta),j(\theta)$ of Sec.~\ref{sec:flux}.
One obtains
\begin{equation}
 \partial_u j=\frac1{\sqrt2}
 (\partial_\theta+\ii\csc\theta\partial_\phi+
 \ii a\sin\theta\partial_u+2\cot\theta)g.
\end{equation}
Substituting the Fourier convention gives Eq.~\eqref{eq:radiative-bianchi}.
Peeling must be applied before imposing a singular phantom rotation: the parameter needed to set the radiative Maxwell scalar to zero generally grows with radius.

\subsection{Energy-flux normalization}

For the flux factors, the asymptotic tetrad is
$\ell=\partial_t+\partial_r$, $n=(\partial_t-\partial_r)/2$, and
$m=(e_\theta+\ii e_\phi)/\sqrt2$.
Direct contraction of a transverse plane wave gives, up to overall signs that drop out of energy densities,
\begin{equation}
 \Phi_2=\frac{E_\theta-\ii E_\phi}{\sqrt2},\qquad
 \Psi_4=\frac12\partial_u^2(h_+-\ii h_\times).
\end{equation}
The Poynting flux of the Maxwell field in Gaussian units is $(E_\theta^2+E_\phi^2)/(4\pi)$, and the gravitational radiation flux is
$[(\partial_u h_+)^2+(\partial_u h_\times)^2]/(16\pi)$.
These yield Eq.~\eqref{eq:flux-normalization} for physical spin-complex amplitudes, with consistent time averaging where needed.
The real and imaginary parts of these amplitudes represent the two polarizations; a single monochromatic spin-complex exponential describes circular polarization, while real linearly polarized combinations require the corresponding consistent time average.
A symmetric tetrad has different boost factors and hence different coefficients when expressed in its NP scalars; those coefficients must not be used with the present amplitudes.

\section{Separated variables and radiative reconstruction}\label{app:physical}

We first specify the NP-to-separated map, then invert it in the charge-regular variables.
The overall mode factors are fixed separately before listing the final normalized coefficients.
Although the intermediate phantom-gauge variables are singular at zero charge, the normalized physical combinations have a regular Kerr limit.

\subsection{From NP fields to separated variables}\label{app:NP-to-separated}

For $Q\ne0$, the sequence is NP fields $\to y_\pm(r,\theta)\to z_\pm(r,\theta)$, followed by separation into angular functions $f_1,f_2$ and radial functions $w,w_\sharp$.
We specify each step following Steps 1--3 of Ref.~\cite{Hintz:2026jiw}.
Besides Eq.~\eqref{eq:source-NP-variables}, it uses the rescaled perturbed spin coefficients
\begin{equation}
 \mathfrak k=\frac{\delta\kappa_{\rm NP}}{\sqrt2(\varrho^*)^2},\qquad
 \mathfrak s=\frac{\varrho\,\delta\sigma_{\rm NP}}{(\varrho^*)^2},\qquad
 \mathfrak l=\frac{\varrho^*\delta\lambda_{\rm NP}}2,\qquad
 \mathfrak n=\frac{\varrho\varrho^*\delta\nu_{\rm NP}}{\sqrt2}.
\end{equation}
The chiral fields, obtained by diagonalizing the derivative terms, are
\begin{equation}
\begin{aligned}
 y_\pm^{(+)}&=\Delta_Q^{5/4}\sqrt{\sin\theta}
 \begin{pmatrix}\psi_2\pm\ii\Delta_Q^{-1/2}\psi_1\\-2Q(\mathfrak k\pm\ii\Delta_Q^{-1/2}\mathfrak s)\end{pmatrix},\\
 y_\pm^{(-)}&=\Delta_Q^{-3/4}\sqrt{\sin\theta}
 \begin{pmatrix}\psi_{-2}\mp\ii\Delta_Q^{1/2}\psi_{-1}\\2Q(\mathfrak n\mp\ii\Delta_Q^{1/2}\mathfrak l)\end{pmatrix}.
\end{aligned}
\label{eq:full-chiral-map}
\end{equation}

The integrating-factor transformation is given by
\begin{equation}
 z_+=(q-M^2)X^{-1}y_+,\qquad z_-=X^*y_-,
 \label{eq:forward-integrating-map}
\end{equation}
where the integrating factor is
\begin{equation}
 X=\begin{pmatrix}
 Q(\sqrt{\Delta_Q}+A_\theta)/\varrho&-\ii C_\varrho\\
 \ii(M-q/\varrho^*)&Q(\sqrt{\Delta_Q}-A_\theta)/\varrho^*
 \end{pmatrix},\qquad \det X=q-M^2.\label{eq:integrating-factor}
\end{equation}
and $X^*$ denotes the Hermitian transpose for real background parameters.
To separate the transformed system, choose an angular eigenfunction $\boldsymbol f(\theta)=(f_1(\theta),f_2(\theta))^T$ satisfying Eq.~\eqref{eq:restricted-angular} with eigenvalue $\lambda$.
Its embedding $\boldsymbol u=(\ii f_1,f_1,-\ii f_2,f_2)^T$ lies in the $\mathcal S=\ii$ eigenspace of the matrix in Eq.~\eqref{eq:four-angular}; the paired $\mathcal S=-\ii$ sector is fixed as described in Sec.~\ref{sec:review}.
Introduce radial functions $z_1(r),z_2(r)$ and set $p(\theta)=(f_1+f_2)/\sqrt2$, $n(\theta)=(f_1-f_2)/\sqrt2$.
For either spin sign, the separation ansatz is
\begin{equation}
\begin{aligned}
 z_+&=\sqrt2(\ii z_1n+z_2p,\ z_1p-\ii z_2n)^T,\\
 z_-&=\sqrt2(\ii z_1n-z_2p,\ z_1p+\ii z_2n)^T.
\end{aligned}
\label{eq:transformed-separation}
\end{equation}

The radial factors defining $w,w_\sharp$ are
\begin{equation}
\begin{aligned}
 (z_1,z_2)&=\begin{cases}
 H_Q(\Delta_Q^{-1/4}w_\sharp,\Delta_Q^{-3/4}w),&\sigma=+,\\
 H_Q(\Delta_Q^{3/4}w,\Delta_Q^{5/4}w_\sharp),&\sigma=-,
 \end{cases}
\end{aligned}
\label{eq:radial-factor-map}
\end{equation}
Together, Eqs.~\eqref{eq:full-chiral-map}--\eqref{eq:radial-factor-map} specify how the angular functions $f_1,f_2$ and radial functions $w,w_\sharp$ encode the NP fields.

\subsection{From regular separated variables to radiative fields}\label{app:regular-to-fields}

We now start with $(\chi_\epsilon,\zeta_\epsilon,w,w_\flat)$, undo the regularizing rescalings, and then invert the transformations above.
The intermediate expressions use $Q\ne0$; the final normalized combinations extend to $Q=0$.
For the charge-regular variables of Secs.~\ref{sec:angular} and \ref{sec:radial}, the inverse rescalings are
\begin{align}
 (n,p)&=\begin{cases}(\chi,\ii Q\zeta/(3M)),&\epsilon=+,\\
 (-\ii Q\zeta/(3M),\chi),&\epsilon=-,
 \end{cases}\\
 w_\sharp&=\begin{cases}Qw_\flat/(3\ii M^2),&\sigma\epsilon=+,\\
 3\ii w_\flat/Q,&\sigma\epsilon=-.
 \end{cases}
\end{align}
Substitute these rescalings into Eqs.~\eqref{eq:radial-factor-map} and \eqref{eq:transformed-separation} to obtain $z_\pm$.
The inverse integrating-factor relations are $y_+=Xz_+/(q-M^2)$ and $y_-=(X^*)^{-1}z_-$, with the same $X$ as above.
For clarity, the inverse needed in the second relation is
\begin{equation}
 (X^*)^{-1}=\frac1{q-M^2}
 \begin{pmatrix}
 Q(\sqrt{\Delta_Q}-A_\theta)/\varrho&\ii C_\varrho\\
 -\ii(M-q/\varrho^*)&Q(\sqrt{\Delta_Q}+A_\theta)/\varrho^*
 \end{pmatrix}.
\end{equation}
Using the explicit $X$ in Eq.~\eqref{eq:integrating-factor}, their first components obey
\begin{align}
 (y_++y_-)_1&=\frac{2\sqrt2}{q-M^2}
 \left[\frac{\ii Q\sqrt{\Delta_Q}}{\varrho}z_1n+
 \frac{QA_\theta}{\varrho}z_2p-C_\varrho z_2n\right],\\
 (y_+-y_-)_1&=\frac{2\sqrt2}{q-M^2}
 \left[\frac{\ii QA_\theta}{\varrho}z_1n+
 \frac{Q\sqrt{\Delta_Q}}{\varrho}z_2p-\ii C_\varrho z_1p\right].
\end{align}
To reconstruct the Weyl fields, take the sum and difference of the first components of Eq.~\eqref{eq:full-chiral-map}:
\begin{equation}
\begin{aligned}
 (y_+^{(+)}+y_-^{(+)})_1&=2\Delta_Q^{5/4}\sqrt{\sin\theta}\,\psi_2,&
 (y_+^{(+)}-y_-^{(+)})_1&=2\ii\Delta_Q^{3/4}\sqrt{\sin\theta}\,\psi_1,\\
 (y_+^{(-)}+y_-^{(-)})_1&=2\Delta_Q^{-3/4}\sqrt{\sin\theta}\,\psi_{-2},&
 (y_+^{(-)}-y_-^{(-)})_1&=-2\ii\Delta_Q^{-1/4}\sqrt{\sin\theta}\,\psi_{-1}.
\end{aligned}
\label{eq:chiral-radiative}
\end{equation}

\subsection{Overall mode normalization}\label{app:mode-normalization}

The phantom-gauge NP normalization is fixed in Eq.~\eqref{eq:source-NP-variables}.
To fix the mode convention, substitute the inverse rescalings above and compare the surviving raw radiative field with $R_sS_s$ as $Q\to0$.
The second row below gives the leading multiplicative factors, including their displayed powers of $Q$; a raw field may therefore diverge or vanish in this comparison.
We use their reciprocals at every $Q$, with no charge dependence beyond the explicit power shown, to normalize the complete phantom-gauge mode:
\begin{equation}
\begin{array}{c|cccc}
(\sigma,\epsilon)&(+,+)&(+,-)&(-,+)&(-,-)\\\hline
\text{leading raw factor}&\sqrt2/M^5&\ii/M^4&3\ii\sqrt2M^3/Q&-2Q/3
\end{array}.
\label{eq:mode-normalization}
\end{equation}
The composition with the inverse variable transformations is regular at $Q=0$ even when an individual factor is not.

For example, take $(\sigma,\epsilon)=(-,-)$, so that $s=-1$, $n=-\ii Q\zeta/(3M)$, $p=\chi$, and $w_\sharp=Qw_\flat/(3\ii M^2)$.
Equation~\eqref{eq:radial-factor-map} then gives $z_1=H_Q\Delta_Q^{3/4}w$ and $z_2=H_Q\Delta_Q^{5/4}Qw_\flat/(3\ii M^2)$.
Insert these expressions into $(y_+-y_-)_1$ above and divide by $-2\ii\Delta_Q^{-1/4}\sqrt{\sin\theta}$ in Eq.~\eqref{eq:chiral-radiative} to obtain $\psi_{-1}$.
Equation~\eqref{eq:phantom-radiative-fields} gives $\mathcal E_-^{\rm raw}=\sqrt2Q\,\psi_{-1}^{\rm raw}/(3M)$, where $\psi_{-1}^{\rm raw}$ is the just reconstructed NP variable before the overall mode factor is applied.
Thus,
\begin{equation}
 \mathcal E_-^{\rm raw}
 =-\frac{2QH_Q\Delta_Q}{3M^2(M^2-q)\sqrt{\sin\theta}}
 \left[MC_\varrho w\chi+\frac{q}{3\varrho}
 \left(\frac{\Delta_Q}{M}w_\flat\chi+\ii A_\theta w\zeta\right)\right].
 \label{eq:worked-reconstruction}
\end{equation}
Its leading ratio to $R_{-1}S_{-1}$ is $-2Q/3$, fixing the last entry of Eq.~\eqref{eq:mode-normalization}.
After multiplying by $-3/(2Q)$, the prefactor is precisely $\mathcal B_{-1}\Delta_Q$, and the bracket is $e_{--}$ below.
The other three branches follow by the same substitutions into the sum or difference selected in Eq.~\eqref{eq:chiral-radiative}.

\subsection{Normalized reconstruction coefficients}\label{app:reconstruction-coefficients}

After factoring out $\mathcal B_s\Delta_Q^{-2\sigma}$ or $\mathcal B_s\Delta_Q^{-\sigma}$, the final reconstruction~\eqref{eq:normalized-radiative-fields} has coefficients
\begin{align}
 g_{++}&=MC_\varrho w\chi-\frac{q}{3\varrho}
       \left(\frac{\Delta_Q}{M}w_\flat\chi+\ii A_\theta w\zeta\right),\notag\\
 e_{++}&=\frac{Q^3}{27\sqrt2M^3\varrho}
       \left(\varrho C_\varrho w_\flat\zeta+3\ii MA_\theta w_\flat\chi-3M^2w\zeta\right),\notag\\
 g_{+-}&=\frac{\sqrt2MQ}{3\varrho}
       \left(-3\Delta_Qw_\flat\zeta+3\ii MA_\theta w\chi-\varrho C_\varrho w\zeta\right),\notag\\
 e_{+-}&=MC_\varrho w_\flat\chi+\frac{q}{3\varrho}
       (Mw\chi+\ii A_\theta w_\flat\zeta),\notag\\
 g_{-+}&=MC_\varrho w_\flat\chi-\frac{q}{3\varrho}
       (Mw\chi+\ii A_\theta w_\flat\zeta),\notag\\
 e_{-+}&=\frac{\sqrt2Q^3}{27M\varrho}
       \left(3\Delta_Qw_\flat\zeta-3\ii MA_\theta w\chi-\varrho C_\varrho w\zeta\right),\notag\\
 g_{--}&=\frac{Q}{3\sqrt2M\varrho}
       \left(\varrho C_\varrho w_\flat\zeta-3\ii MA_\theta w_\flat\chi+3M^2w\zeta\right),\notag\\
 e_{--}&=MC_\varrho w\chi+\frac{q}{3\varrho}
       \left(\frac{\Delta_Q}{M}w_\flat\chi+\ii A_\theta w\zeta\right).
\label{eq:reconstruction-coefficients}
\end{align}
Here, $\chi=\chi_\epsilon$ and $\zeta=\zeta_\epsilon$ in the corresponding row.
These are coefficients of the exact fields~\eqref{eq:normalized-radiative-fields}, not their asymptotic radiation amplitudes.
Their relation to the latter is the large-radius extraction in Eq.~\eqref{eq:reconstruction-to-radiation}.

At fixed exterior radius and interior angle, their $Q=0$ values follow by $C_\varrho\to M$, $H_Q\to H_0$, and $M^2-q\to M^2$.
Specifically, the nonzero entries at $Q=0$ are
\begin{equation}
\begin{aligned}
 g_{++}(0)&=M^2w\chi_+,&g_{-+}(0)&=M^2w_\flat\chi_+,\\
 e_{+-}(0)&=M^2w_\flat\chi_-,&e_{--}(0)&=M^2w\chi_-.
\end{aligned}
\end{equation}
All other entries vanish.
They give Eq.~\eqref{eq:unit-kerr-field} without a singular change of tetrad at $Q=0$.

\section{Angular eigenbranches and boundary normalization}\label{app:boundary-proof}

This appendix establishes the parameter dependence used in Secs.~\ref{sec:angular-branch} and \ref{sec:scope}.
We first impose both angular endpoint conditions and obtain a local separation curve $\mu(q)$.
Given such a curve, we then control radial solutions on a finite interval, fix their infinity and horizon normalizations, and form the connection matrix.
The angular argument holds at fixed real $c=a\omega$; the boundary-normalization argument additionally assumes real $\omega\ne0$ and horizon nonresonance.
The finite-interval radial statement below has weaker hypotheses and is stated separately.

\subsection{Angular eigenbranches}\label{app:angular-proof}

Fix $m$, $\sigma$, $\epsilon$, real $c$, and a simple Kerr eigenvalue $\mu_0$ as in Sec.~\ref{sec:angular-branch}.
To prove the local analytic continuation of this regular angular eigenmode, we take three steps: keep the operator domain fixed, solve the nonlinear eigenvalue condition, and verify that the inverse map satisfies the first-order endpoint conditions.

\subsubsection{A fixed domain and bounded perturbation}

To apply a bounded-perturbation argument, we first remove the first derivative in Eq.~\eqref{eq:angular-scalar}:
\begin{equation}
 \chi_\epsilon=\sqrt{\beta_Q}\,z.
 \label{eq:angular-branch-map}
\end{equation}
The square root of $\beta_Q$ is chosen to approach one as $Q\to0$.
Where $\beta_Q\ne0$, this is an invertible change of the dependent variable in Eq.~\eqref{eq:explicit-teuk-angular}, not an additional eigenvalue equation.
The second-order equation~\eqref{eq:angular-scalar} becomes
\begin{equation}
\mathsf L_\theta(q,\mu)z=\mu z,
\end{equation}
where
\begin{equation}
 \begin{split}
 \mathsf L_\theta(q,\mu)&=-\partial_\theta^2+U(q,\mu),\\
 U&=t_Q^2-\epsilon t_Q'+\epsilon q_Qt_Q+3Mb_Q+q b_Q^2-\frac12q_Q'+\frac14q_Q^2.
 \end{split}
 \label{eq:branch-operator}
\end{equation}

The singular terms of this potential are
\begin{equation}
 \begin{aligned}
 U(q,\mu)&=\frac{(m+s)^2-1/4}{\theta^2}+\order{1}, &&\theta\to0,\\
 U(q,\mu)&=\frac{(m-s)^2-1/4}{(\pi-\theta)^2}+\order{1}, &&\theta\to\pi.
 \end{aligned}
\end{equation}
Their coefficients are independent of $q$, so the singular part can be kept fixed.
The difference $U(q,\mu)-U(0,\mu)$ is bounded on the whole interval and holomorphic in $(q,\mu)$ near $(0,\mu_0)$.
Indeed, $t_Q-t_0$ is proportional to $q\sin\theta$ as in \eqref{eq:tQ0bQ0}, $q_Q=\beta_Q^\prime/\beta_Q$ is proportional to $q\sin\theta/\beta_Q$, and $\beta_Q$ stays uniformly away from zero.
Products with the singular $t_0$ are bounded because $t_0\sin\theta$ is a polynomial in $\cos\theta$.
All remaining differences and derivatives are bounded as well.
Define the Kerr operator $\mathsf L_{\theta,0}=\mathsf L_\theta(0,\mu)$, which is independent of $\mu$.
Its regular eigenfunctions have endpoint behavior $z\sim\theta^{1/2+|m+s|}$ and $z\sim(\pi-\theta)^{1/2+|m-s|}$; when $m\pm s=0$, the logarithmic solution is excluded.
We keep this regular operator domain in $L^2(0,\pi;\dd\theta)$ as the charge varies.
The Kerr operator is self-adjoint on this domain and has discrete, simple fixed-$m$ eigenvalues.

\subsubsection{Spectral projection and self-consistency}

At nonzero charge, the potential itself depends on the unknown separation parameter.
To separate this dependence from the spectral problem, first insert an independent trial value $\widehat\mu$ into $\mathsf L_\theta(q,\widehat\mu)$ and compute its eigenvalue near $\mu_0$.
Call that eigenvalue $\mathscr E(q,\widehat\mu)$.
It need not equal the value $\widehat\mu$ inserted in the coefficients; the physical separation parameter is obtained only after imposing $\mathscr E(q,\widehat\mu)=\widehat\mu$.

To isolate the chosen Kerr mode, choose a positively oriented circle $\Gamma$ enclosing $\mu_0$ and no other eigenvalue of $\mathsf L_{\theta,0}$.
Let $\eta$ be the auxiliary complex spectral coordinate on this circle.
The resolvent
\begin{equation}
 \mathscr R_0(\eta)=(\eta-\mathsf L_{\theta,0})^{-1}
\end{equation}
maps a prescribed forcing to the solution of the inhomogeneous equation with operator $\eta-\mathsf L_{\theta,0}$.
Here and below, operator norms are those on $L^2(0,\pi;\dd\theta)$.
The spectral theorem bounds $\|\mathscr R_0(\eta)\|$ by the inverse distance from $\eta$ to the Kerr spectrum, uniformly on $\Gamma$.
The perturbation $\mathsf V=\mathsf L_\theta(q,\widehat\mu)-\mathsf L_{\theta,0}$ is multiplication by the bounded potential difference, so its operator norm equals the supremum norm of that difference and tends to zero with $q$, uniformly for $\widehat\mu$ near $\mu_0$.
Consequently, $\|\mathsf V\mathscr R_0\|<1$ for sufficiently small $q$, and
\begin{equation}
 (\eta-\mathsf L_\theta(q,\widehat\mu))^{-1}
 =\mathscr R_0(\eta)\sum_{n=0}^\infty
 [\mathsf V\mathscr R_0(\eta)]^n.
\end{equation}
This convergent operator geometric series, called the Neumann series, gives holomorphic dependence on $(q,\widehat\mu)$.
The contour integral
\begin{equation}
 \Pi(q,\widehat\mu)=\frac1{2\pi\ii}\oint_\Gamma
 (\eta-\mathsf L_\theta(q,\widehat\mu))^{-1}\,\dd\eta
\end{equation}
defines a spectral projection, $\Pi^2=\Pi$.
At zero charge it projects onto the chosen Kerr mode; continuity preserves its rank one for small $q$.
There is therefore exactly one enclosed eigenvalue, $\mathscr E(q,\widehat\mu)$, and it remains simple.
The integral is the residue of this simple resolvent pole, and its range is the associated one-dimensional eigenspace.
Both the eigenvalue and this projection depend holomorphically on the parameters.

At $q=0$, every trial value gives exactly the same operator $\mathsf L_{\theta,0}$, hence exactly the same eigenvalue $\mu_0$:
\begin{equation}
 \mathscr E(0,\widehat\mu)=\mu_0,\qquad
 \partial_{\widehat\mu}\mathscr E(0,\widehat\mu)=0.
\end{equation}
Thus, $\partial_{\widehat\mu}(\mathscr E(q,\widehat\mu)-\widehat\mu) = -1$ at $(0,\mu_0)$ and is nonzero.
The analytic implicit-function theorem therefore gives a unique $\widehat\mu=\mu(q)$ satisfying $\mathscr E(q,\mu(q))=\mu(q)$.
This self-consistency condition enforces the original nonlinear eigenvalue equation, rather than merely continuing an eigenvalue of an operator with externally fixed coefficients.
With $c$, $m$ and the other background parameters held real, complex conjugation of its coefficients shows that $\overline{\mu(\bar q)}$ is another holomorphic solution branch with the same initial value $\mu_0$.
Local uniqueness therefore gives $\mu(q)=\overline{\mu(\bar q)}$; in particular, $\mu(q)$ is real for real $q$.
The spectral projection fixes an analytic eigenfunction up to normalization; the real-frequency unit norm may be imposed continuously.

\subsubsection{Reconstruction at the angular poles}

Finally, the inverse~\eqref{eq:angular-reconstruction} must belong to the original first-order domain.
At the north pole, write $\chi\sim\theta^\alpha$, where $\alpha=1/2+|m+s|$ is the regular Frobenius exponent; $\sqrt{\beta_Q}$ does not change this leading power.
The possibly most singular term of $(\partial_\theta+\epsilon t_Q)\chi$ has exponent $\alpha-1$ and coefficient proportional to $\alpha+\epsilon(3/2+\sigma m)$.

It is square integrable unless $m+s=0$; in that case its coefficient vanishes for both entries of Eq.~\eqref{eq:dictionary}, and the next term is of order $\theta^{3/2}$.
Indeed, $\beta_Q$ and the regular part of the effective potential are even in the distance from the pole ($\theta$ at the north pole and $\pi-\theta$ at the south pole), so the nonlogarithmic Frobenius solution has only even corrections to its leading power.
The logarithmic solution would instead recover a nonintegrable $\theta^{-1/2}$ term and is excluded by the regular, or Friedrichs, boundary condition.
The same cancellation occurs at the south pole when $m-s=0$.
Both reconstructed components are therefore square integrable and satisfy $\mathsf A_Q\boldsymbol f=\lambda\boldsymbol f$ in the maximal domain used in Ref.~\cite{Hintz:2026jiw}.
Thus, $\lambda=\epsilon hL(\mu(q))$ is an eigenvalue of the original angular operator for nonzero real $Q$.

This connects the angular eigenvalue branch to the separated angular problem, without asserting a full spacetime reconstruction.
The regular Frobenius expansions have fixed leading powers and coefficients holomorphic in $(q,\mu)$, uniformly on small pole neighborhoods.
Applying the cancelling combination $\zeta_\epsilon=\beta_Q^{-1}\delta_\epsilon\chi_\epsilon$ from Eq.~\eqref{eq:angular-division-free} preserves an integrable leading power there.
Together with analytic dependence on the intervening compact interval, this gives a holomorphic $L^2$ pair $(\chi_\epsilon,\zeta_\epsilon)$ along the branch.
These are the norms needed for the radiation formulas; the uncombined derivative $\chi_\epsilon'$ need not be in $L^2$ at a critical pole.

\subsection{Radial solutions on a finite interval}\label{app:radial-local}

We next establish continuous and analytic charge dependence of radial solutions away from the endpoints.
This finite-interval result will propagate the endpoint-normalized bases to a common matching radius.
Fix a compact interval $J$ strictly outside $r_+(0)$, a finite $\omega$, and a continuous bounded parameter curve $\mu(Q^2)$ with a finite limit.
Let $\mathsf A_r(r,q)$ denote the coefficient matrix displayed in Eq.~\eqref{eq:radial-Y-system}.
It converges uniformly on $J$ as $q\to0$.
Define the initial-value fundamental matrix by
\begin{equation}
 \partial_r\mathsf X(r,q)=\mathsf A_r(r,q)\mathsf X(r,q),\qquad
 \mathsf X(r_b,q)=\Id,\qquad r_b\in J.
 \label{eq:local-radial-fundamental}
\end{equation}
This normalization is at a finite radius, distinct from the horizon and infinity normalizations.
The integral equation
\begin{equation}
 \mathsf X(r,q)=\Id+\int_{r_b}^{r}\mathsf A_r(t,q)\mathsf X(t,q)\,\dd t
\end{equation}
shows that $\mathsf X(r,q)$ converges uniformly to $\mathsf X(r,0)$ on $J$, as follows.
A uniform bound on $\mathsf A_r$ first bounds $\mathsf X$ on $J$.
Subtracting the equations at $q$ and zero and applying Gronwall's inequality then yields
\begin{equation}
 \sup_J\|\mathsf X(r,q)-\mathsf X(r,0)\|
 \le C_J\sup_J\|\mathsf A_r(r,q)-\mathsf A_r(r,0)\|,
\end{equation}
with $C_J$ independent of sufficiently small $q$.
This estimate explains both fixed-interval convergence and why it does not by itself control the infinite exterior.
If $\mu$ is holomorphic in $Q^2$, the same integral equation, iterated on $J$, gives holomorphic local solutions.
This argument applies to any prescribed curve satisfying these hypotheses; Appendix~\ref{app:angular-proof} supplies such curves locally for the regular angular modes.

\subsection{Infinity normalization}\label{app:infinity-proof}

We now fix a real $\omega\ne0$ and assume that $\mu(q)$ is bounded and holomorphic near $q=0$.
We may use the angular curve constructed in Appendix~\ref{app:angular-proof}, or any prescribed curve with the same regularity.
To derive the infinity normalization in Sec.~\ref{sec:boundary-normalization}, write $\mathcal P_\tau=p_1r+p_0$ and $\mathcal R_\tau=\rho_1r+\rho_0$ using Eqs.~\eqref{eq:FG-r-dep} and \eqref{eq:radial-coupling-coefficients-renamed}.
The state transformation~\eqref{eq:infinity-state-map} gives
\begin{equation}
\begin{aligned}
 Z'&=\mathsf B(r)Z,\\
 \mathsf B(r)&=\begin{pmatrix}-\ii\xi_Q&\mathcal P_\tau/r\\r\mathcal R_\tau/\Delta_Q&1/r-B_Q/\Delta_Q-\ii\xi_Q\end{pmatrix}\\
 &=\mathsf B_0+\mathsf B_1/r+\order{r^{-2}},\\
 \mathsf B_0&=\begin{pmatrix}-\ii\xi_Q&p_1\\\rho_1&2\ii\omega+\ii\xi_Q\end{pmatrix},\qquad
 \mathsf B_1=\begin{pmatrix}0&p_0\\\rho_0+2M\rho_1&3\sigma+4\ii M\omega\end{pmatrix}.
\end{aligned}
\label{eq:infinity-matrix}
\end{equation}
The matrices $E_\tau$ in Eq.~\eqref{eq:infinity-diagonalization} obey
\begin{equation}
 E_\tau^{-1}\mathsf B_0E_\tau=\operatorname{diag}(0,2\ii\omega).
\end{equation}
Their determinants are $1-q/M^2$, and the two rates are distinct for $\omega\ne0$.
The diagonal of the next coefficient is
\begin{equation}
 \operatorname{diag}(E_\tau^{-1}\mathsf B_1E_\tau)=(3\sigma,4\ii M\omega).
\end{equation}
These are the power exponents, not eigenvalues of $\mathsf B_1$.
Both the exponential rates and power exponents are independent of charge.

Leading rates and powers do not by themselves fix every subdominant contribution to a column.
To fix the basis, we use asymptotic diagonalization followed by a Volterra integral equation with its endpoint at infinity, a standard strategy in asymptotic ODE theory; see Ref.~\cite{Cotti:2023yor} for a modern treatment, including parameter dependence.
We give the reduction and integrable-remainder estimate explicitly below, since the relative powers of the two columns determine how many orders must be removed in this problem.
Choose
\begin{equation}
 \mathsf S(r)=\Id+\sum_{j=1}^4\frac{\mathsf S_j}{r^j},\qquad
 \operatorname{diag}\mathsf S_j=0,
 \label{eq:finite-normal-form}
\end{equation}
such that the off-diagonal entries of
$\mathsf S^{-1}E_\tau^{-1}\mathsf B E_\tau \mathsf S-\mathsf S^{-1}\mathsf S'$
vanish through order $r^{-4}$.
At each order the unknown off-diagonal coefficient appears in its commutator with $\operatorname{diag}(0,2\ii\omega)$.
Division by the nonzero gap $2\ii\omega$ therefore determines it uniquely from the previous orders.
All $\mathsf S_j$ are holomorphic in $q$.
The remaining diagonal coefficients beyond order $r^{-1}$ are left in the remainder.
With
\begin{equation}
 D_\infty(r)=\operatorname{diag}(d_1,d_2),\qquad
 d_1=(r/M)^{3\sigma},\qquad
 d_2=e^{2\ii\omega r}(r/M)^{4\ii M\omega}.
 \label{eq:infinity-powers}
\end{equation}
write the infinity fundamental matrix as
\begin{equation}
 Z_\infty=E_\tau \mathsf S D_\infty \mathsf U_\infty,\qquad
 \mathsf U_\infty(r)=\Id-\int_r^\infty\mathcal K(t)\mathsf U_\infty(t)\,\dd t.
\label{eq:infinity-volterra}
\end{equation}
Here, the kernel is defined by
\begin{align}
 T&=\mathsf S^{-1}E_\tau^{-1}\mathsf B E_\tau \mathsf S-\mathsf S^{-1}\mathsf S',\\
 \mathcal K&=D_\infty^{-1}TD_\infty-D_\infty^{-1}D_\infty'.
\end{align}
The diagonal entries of $\mathcal K$ are $\order{r^{-2}}$.
Its off-diagonal entries are the $\order{r^{-5}}$ remainder multiplied by the ratio of the two entries of $D_\infty$.
For real $\omega$ those ratios have magnitudes $(r/M)^{\pm3\sigma}$, and hence $\mathcal K=\order{r^{-2}}$ in norm.
The estimate holds uniformly in a sufficiently small complex $q$ disk.
The terminal integral equation thus has a unique solution near infinity, obtained by successive iteration, and that solution is holomorphic in $q$.
Continuation to a fixed exterior matching point preserves this property.
Equation~\eqref{eq:infinity-volterra}, including the specified finite reduction, defines the normalization; leading asymptotic powers alone need not distinguish an admixture of a subdominant solution.
Returning to $Y$ gives exactly the matrix in Eq.~\eqref{eq:infinity-matrix-definition}:
\begin{equation}
 \mathsf I=e^{\ii\xi_Qr}\operatorname{diag}(1,M/r)E_\tau \mathsf S D_\infty \mathsf U_\infty \eqqcolon (Y_1^\mathrm{I},Y_2^\mathrm{I}).
 \label{eq:infinity-basis}
\end{equation}
Here, $D_\infty$ is Eq.~\eqref{eq:infinity-powers}; $\mathsf S$ is fixed by Eq.~\eqref{eq:finite-normal-form}, and $\mathsf U_\infty$ by Eq.~\eqref{eq:infinity-volterra}.
The residual matrix is what remains after removing all preceding factors.
Its identity limit fixes every column, excluding any constant right factor or admixture left invisible by leading asymptotics alone.

For comparison with the horizon columns, the infinity columns have the leading asymptotic form
\begin{equation}
 Y_j^{\mathrm I}=e^{\ii\xi_Qr}\operatorname{diag}(1,M/r)
 e^{\kappa_j r}(r/M)^{\eta_j}
 \left[u_j^{(0)}+\order{M/r}\right],\qquad
 (\kappa_1,\kappa_2)=(0,2\ii\omega),\quad(\eta_1,\eta_2)=(3\sigma,4\ii M\omega).
 \label{eq:infinity-column-expansion}
\end{equation}
Here, $u_j^{(0)}$ is column $j$ of $E_\tau$.
These leading vectors give the coefficients $A_\tau(q)$ and $B_\tau(q)$ in Eq.~\eqref{eq:finite-infinity-radial} after applying the radial map and $D_\mathrm I^{-1}$.
The exact columns remain fixed by the terminal prescription~\eqref{eq:infinity-volterra}; leading asymptotics alone do not specify their subdominant admixtures.

\subsubsection{Kerr normalization at infinity}\label{app:kerr-infinity-normalization}

We now determine the factors that convert the infinity columns into unit-amplitude Kerr radial waves.
For a state $Y=(w,w_\flat)^T$, we write $W_\tau[Y]\coloneqq e_\tau^T Y$ for its retained component: $W_+[Y]=w$ and $W_-[Y]=w_\flat$.
At $q=0$, the leading vectors in Eq.~\eqref{eq:infinity-diagonalization} directly give $W_+[Y_1^{\mathrm I}]\sim\rho^{3\sigma}$ and $W_-[Y_2^{\mathrm I}]\sim\rho^{-1}e^{2\ii\omega r}\rho^{4\ii M\omega}$, where $\rho=r/M$.
These fix the unit entries of $D_\mathrm I$.
For the other two combinations the selected component vanishes in the leading vector, accounting for $B_+(0)=A_-(0)=0$.
Their first nonzero terms are needed only at $q=0$ and, as we shall show below, follow directly from the regular first-order equations, without computing the general finite-charge correction.

For $\tau=+1$, the second infinity column has $w_\flat\sim\rho^{-1}e^{2\ii\omega r}\rho^{4\ii M\omega}$.
Since $F_Q=1$ at zero charge, the first equation in Eq.~\eqref{eq:regular-radial-pair} is $Mw'=w_\flat$.
Differentiating the exponential supplies the leading factor $2\ii\omega$, so $w\sim w_\flat/(2\ii M\omega)$.
For $\tau=-1$, the first infinity column instead has $w\sim\rho^{3\sigma}$ and satisfies $\Delta_0w_\flat'+B_0w_\flat=Mw$, where $B_0=B_Q|_{q=0}$.
Here, $B_0=-2\ii\omega r^2+\order{r}$, and the derivative term is one power of $r$ smaller than $B_0w_\flat$ for this nonoscillatory column.
Thus, $w_\flat\sim-Mw/(2\ii\omega r^2)$.
The two retained components are therefore
\begin{equation}
 W_+[Y_2^{\mathrm I}]\sim\frac{\rho^{-1}}{2\ii M\omega}e^{2\ii\omega r}\rho^{4\ii M\omega},\qquad
 W_-[Y_1^{\mathrm I}]\sim-\frac{\rho^{3\sigma-2}}{2\ii M\omega}.
\label{eq:kerr-subleading-components}
\end{equation}
Their coefficients give the two nonunit entries of $D_\mathrm I$ in Eq.~\eqref{eq:boundary-normalization-factors}.

To recover the radial-wave amplitudes, multiply each retained component by the common Kerr radial prefactor,
\begin{equation}
 H_0(\Delta_0/M^2)^{-s}\sim\rho^{-2s}e^{-\ii\omega r}\rho^{-2\ii M\omega}.
\end{equation}
Using $2s=3\sigma+\tau$ and dividing by the corresponding entry of $D_\mathrm I$ gives $\rho^{-1}e^{-\ii\omega r_*}$ for the first column and $\rho^{-2s-1}e^{\ii\omega r_*}$ for the second, each with unit coefficient.
This proves Eq.~\eqref{eq:unit-kerr-boundary} and, in the expansion~\eqref{eq:finite-infinity-radial}, $A_-^{(1)}(0)=B_+^{(1)}(0)=1$.

\subsection{Horizon normalization}\label{app:horizon-proof}

At the horizon, use $d,x$ from Eq.~\eqref{eq:horizon-coordinates} and $\nu_\pm$ from Eq.~\eqref{eq:horizon-exponents}.
The residue matrix of the equation for $Y$ in $x$ is \begin{equation}
 J_\mathrm{H}=\begin{pmatrix}0&0\\M\mathcal R_\tau(r_+)/d&\nu_+\end{pmatrix}.
\end{equation}
Assume $\nu_+\notin\mathbb Z$ in the parameter neighborhood under consideration.
For real frequency this excludes the threshold $K_0(r_+)=0$ and guarantees nonresonance.
The Frobenius solutions normalized by
\begin{equation}
 \mathsf H=(Y^\mathrm{H}_1,Y^\mathrm{H}_2),\qquad
 Y^\mathrm{H}_1=\begin{pmatrix}1\\M\mathcal R_\tau(r_+)/B_Q(r_+)\end{pmatrix}+\order{x},\qquad
 Y^\mathrm{H}_2=x^{\nu_+}\left[\begin{pmatrix}0\\1\end{pmatrix}+\order{x}\right],
\label{eq:horizon-basis}
\end{equation}
are holomorphic in $q$ at any fixed exterior point.
The logarithm is chosen real for positive $x$ at real parameters and continued locally in $q$.
This is a normalization of the separated master variables, not yet a normalization to unit physical energy flux.

The diagonal factor $D_\mathrm H$ in the main text follows by comparing the desired waves in Eq.~\eqref{eq:finite-horizon-radial} with the retained components of these Frobenius solutions:
\begin{equation}
\begin{array}{c|cc}
 &W_\tau[Y_1^\mathrm{H}]&W_\tau[Y_2^\mathrm{H}]\\\hline
\tau=+1&1+\order{x}&\dfrac{F_+}{\nu_++1}x^{\nu_++1}[1+\order{x}]\\
\tau=-1&\dfrac{MF_+}{B_+}+\order{x}&x^{\nu_+}[1+\order{x}]
\end{array}.
\label{eq:horizon-retained-components}
\end{equation}
The radial prefactor has leading coefficient $e^{-\ii(\omega+\xi_Q)r_+}(d/M)^{-s-\ii\alpha_-}$.
The desired ingoing and outgoing waves have coefficients $(d/M)^{-s}e^{-\ii k_\mathrm H r_*^{\mathrm H}}$ and $e^{\ii k_\mathrm H r_*^{\mathrm H}}$, respectively.
Dividing the retained coefficients, including the common prefactor, by these desired coefficients gives $D_\mathrm H$ in Eq.~\eqref{eq:boundary-normalization-factors}.

\subsection{Connection matrix and normalization check}\label{app:connection-proof}

The two normalized bases~\eqref{eq:infinity-basis} and \eqref{eq:horizon-basis} can now be compared at any common exterior radius $r_b$.
Their analytic dependence and invertibility imply that $C(q)=\mathsf I(r_b,q)^{-1}\mathsf H(r_b,q)$ is holomorphic near zero charge.
Because both bases solve the same first-order system, this matrix is independent of $r_b$.
An exact determinant identity checks the relative normalization and makes the nonzero Kerr limit explicit.
Abel's identity applies to either fundamental matrix:
\begin{equation}
 (\det\mathsf H)'=-\frac{B_Q}{\Delta_Q}\det\mathsf H,\qquad
 (\det\mathsf I)'=-\frac{B_Q}{\Delta_Q}\det\mathsf I.
\end{equation}
Fixing the constants with the respective endpoint normalizations gives
\begin{equation}
\begin{aligned}
 \det\mathsf H&=e^{2\ii(\omega+\xi_Q)(r-r_+)}\left(\frac{r-r_+}{M}\right)^{\nu_+}
 \left(\frac{r-r_-}{d}\right)^{\nu_-},\\
 \det\mathsf I&=(1-q/M^2)e^{2\ii(\omega+\xi_Q)r}
 \left(\frac{r-r_+}{M}\right)^{\nu_+}
 \left(\frac{r-r_-}{M}\right)^{\nu_-},\\
 \det C&=\frac{e^{-2\ii(\omega+\xi_Q)r_+}}{1-q/M^2}
 \left(\frac{M}{d}\right)^{\nu_-},
\end{aligned}
\label{eq:determinant-proof}
\end{equation}
where we used $\nu_++\nu_-=3\sigma-1+4\ii M\omega$ and the asymptotic determinant of
$e^{\ii\xi_Qr}\operatorname{diag}(1,M/r)E_\tau \mathsf S D_\infty \mathsf U_\infty$ to fix the infinity normalization.
The connection determinant $\det C$ has a finite, nonzero Kerr limit.
Ratios of individual connection coefficients require separate attention at zeros of their denominators; invertibility of the full matrix does not exclude such zeros.

We checked this connection construction numerically at $a/M=0.4$, $m=2$, $M\omega=0.7$, and prescribed $\mu=4.2$ for all four branches and $Q/M=0,0.05,0.1$.
Specifically, we integrated the horizon and infinity fundamental matrices separately to the common radius $r_b/M=4$, formed $C=\mathsf I(r_b)^{-1}\mathsf H(r_b)$, and tested its determinant against Eq.~\eqref{eq:connection-determinant} and its entries for numerical convergence.
We used fixed-step fourth-order Runge--Kutta integration, starting at $(r-r_+)/M=0.05$ and $r/M=40$ with order-16 endpoint series and 4000 and 16000 steps, respectively.
The relative residual in Eq.~\eqref{eq:connection-determinant} was below $4.7\times10^{-9}$.
Halving the integration step and varying the infinity starting radius and truncation order changed the connection matrices by less than $1.6\times10^{-8}$ in relative Frobenius norm.
These tests use a prescribed $\mu$, not an imposed angular eigenvalue, and check the master-system connection rather than a physical scattering cross section.

\section{Derivation of the outgoing radiation amplitudes}\label{app:outgoing-profiles}

We derive Eq.~\eqref{eq:physical-outgoing-amplitudes} in two stages: extract the outgoing coefficients of the normalized invariants, then remove the gravitational contribution from the electromagnetic one.
The radial extraction requires a subleading coefficient because the leading terms of the electromagnetic inverse map cancel.
The final simplification uses the angular first-order equations, not a charge expansion.
Throughout this appendix, $\sigma=-1$, $\omega$ is real and nonzero, $D=M^2-q$, and $A_\theta=a\sin\theta$.
The phase convention~\eqref{eq:phase-at-infinity} has unit asymptotic coefficient, and the fields already include the mode normalization in Sec.~\ref{sec:physical}.

\subsection{Subleading radial coefficient}

We first determine the radial information needed to extract the radiation from the normalized field inverse.
The leading radial terms cancel in the mixed electromagnetic invariant, so its first nonzero asymptotic coefficient requires one further order of the outgoing radial solution at finite charge.
Only the part of that correction that survives in the field inverse is needed.
For $\sigma=-1$, write the second infinity column as
\begin{equation}
 Z=e^{2\ii\omega r}(r/M)^{4\ii M\omega}
 [u_0+u_1/r+\order{r^{-2}}].
\end{equation}
The leading vectors are $u_0=(3,1)^T$ for $\epsilon=+$ and $u_0=(-q/(3M^2),1)^T$ for $\epsilon=-$.
The first recurrence is
\begin{equation}
 (\mathsf B_0-2\ii\omega\Id)u_1
 =(4\ii M\omega\Id-\mathsf B_1)u_0.
\end{equation}
Its compatibility uses $L^2=1+4q\mu/(9M^2)$.
A particular solution with second component zero has first component
\begin{equation}
 (u_1)_1=\begin{cases}
 -6M+3q/M-\ii D\mu/[M^2\omega(1+L)],&\epsilon=+,\\
 -\ii D(1+L)/(4M^2\omega)+q(2M^2-q)/(3M^3),&\epsilon=-.
 \end{cases}
\end{equation}
At this order an arbitrary multiple of $u_0$ may be added; the next recurrence fixes it in the fully normalized series.
The leading physical coefficients below do not depend on that freedom.

Insert this expansion into the exact $g_{-\epsilon},e_{-\epsilon}$ of Appendix~\ref{app:physical}, retaining the corrections in $\Delta_Q$ and $\varrho=r+\ii a\cos\theta$ as well.
In both branches the leading eigenvector makes the nominal constant term of $e_{-\epsilon}$ vanish.
The first nonzero term comes from the $r^{-1}$ correction, so it must be retained before taking the limit in Eq.~\eqref{eq:reconstruction-to-radiation}.
The coefficients in Eq.~\eqref{eq:reconstruction-coefficients} include the mode normalization; no further factor is applied.
For example, in the $\epsilon=-$ branch, suppress the common outgoing phase and power.
Then $w=-q/(3M^2)+\order{r^{-1}}$, $w_\flat=M/r+\order{r^{-2}}$, and the two constant terms in $e_{--}$ cancel:
\begin{equation}
 M C_\varrho w\chi_-+
 \frac{q\Delta_Q}{3M\varrho}w_\flat\chi_-
 =\left(-\frac q3+\frac q3\right)\chi_-+\order{r^{-1}}.
\end{equation}
The remaining term proportional to $A_\theta$ is also of order $r^{-1}$.
Adding a multiple of $u_0$ to $u_1$ contributes the same vanishing leading combination at this order, which explains the normalization independence of the extracted coefficient.

\subsection{Profiles of the mixed invariants}

We now use the radial expansion to obtain the angular profiles of the two normalized invariants at infinity.
The coefficient $g(\theta)$ is already the physical gravitational radiation amplitude, whereas $e(\theta)$ still contains the gravitational contribution in Eq.~\eqref{eq:asymptotic-mixed}.
Its removal is the task of the next subsection.
With a master outgoing coefficient $A_{\rm out}$, substitution into Eq.~\eqref{eq:reconstruction-to-radiation} gives
\begin{equation}
 g(\theta)=A_{\rm out}\gamma_\epsilon,\qquad
 e(\theta)=A_{\rm out}\eta_\epsilon.
\end{equation}
The exact profiles are
\begin{equation}
\begin{aligned}
 \sqrt{\sin\theta}\,\gamma_+&=\frac{\chi_+}{M^3},\\
 \sqrt{\sin\theta}\,\gamma_-&=\frac{Q}{3\sqrt2 M^3}\zeta_-,\\
 \sqrt{\sin\theta}\,\eta_+
 &=\frac{\ii\sqrt2 Q^3}{9M^4D}\left[-3A_\theta\chi_+-a\cos\theta\,\zeta_+ +\frac{D\mu \zeta_+}{3M^2\omega(1+L)}\right],\\
 \sqrt{\sin\theta}\,\eta_-
 &=-\frac{\ii}{M^2}\left[\frac{1+L}{4\omega}\chi_- +\frac{aq\cos\theta}{3D}\chi_-
 +\frac{A_\theta q^2}{9M^2D}\zeta_-\right].
\end{aligned}
\label{eq:outgoing-profiles}
\end{equation}
Here, $\chi_\epsilon,\zeta_\epsilon$ denote the two angular branches at $\sigma=-1$.
No charge expansion is made here.
The leading infinity eigenvectors suffice for $\gamma_\epsilon$, whereas the next radial order is essential for $\eta_\epsilon$.
The arbitrary multiple of $u_0$ in the first recurrence cancels from both extracted profiles.

\subsection{Extraction of the physical Maxwell amplitude}

Finally, we turn the mixed coefficient $e(\theta)$ into the physical Maxwell amplitude $f(\theta)$ using Eq.~\eqref{eq:physical-Maxwell-extraction}.
The gravitational subtraction differentiates $g(\theta)$ with respect to angle; the angular first-order equations express this derivative in the same components as the profiles above, making their cancellation explicit.
For either angular component $v$, at $\sigma=-1$,
\begin{equation}
 \sqrt{\sin\theta}\,\mathscr L_2^\dagger
 \frac{v}{\sqrt{\sin\theta}}
 =(\partial_\theta+t_Q)v-\frac{2\omega q A_\theta}{D}v.
 \label{eq:radiation-angular-reduction}
\end{equation}
The first-order angular equations give $(\partial_\theta+t_Q)\chi_+=\beta_Q\zeta_+$ and $(\partial_\theta+t_Q)\zeta_-=\gamma_Q\chi_-$.
Inserting these relations into Eq.~\eqref{eq:physical-Maxwell-extraction}, all explicit $A_\theta$ and $a\cos\theta$ terms cancel those in $\eta_\epsilon$.
Using $4q\mu=9M^2(L^2-1)$, the remaining coefficients reduce to $Q\zeta_+/(3\sqrt2M^4\omega)$ and $\chi_-/(2M^2\omega)$, with the common factor $-\ii A_{\rm out}/\sqrt{\sin\theta}$.
Together with the gravitational profiles above, they give exactly Eq.~\eqref{eq:physical-outgoing-amplitudes}.
Thus, the apparently higher-order mixed profile on the gravitational branch does not suppress the order-$Q/M$ physical Maxwell radiation.

\end{widetext}

\bibliography{ref}
\end{document}